\documentclass[a4paper,pre,reqno,superscriptaddress,twocolumn,floatfix,longbibliography]{revtex4-1}
\usepackage{graphicx,color}
\usepackage{dcolumn}
\usepackage{epsfig}  
\usepackage[centertags]{amsmath}
\usepackage{amsfonts}
\usepackage{euscript}
\usepackage{amssymb}
\usepackage{amsthm}
\usepackage{newlfont}
\usepackage{lipsum}
\usepackage{mathtools}
\usepackage{mathrsfs}
\usepackage{subfigure}
\usepackage{todonotes}
\usepackage[normalem]{ulem}
\usepackage[colorlinks=true,linkcolor=blue,urlcolor=blue,citecolor=blue,bookmarks=true]{hyperref}
\usepackage{siunitx} 

\usepackage{multirow}
\usepackage{longtable}
\usepackage{tabularray}
\usepackage{booktabs}

\usepackage{comment}
\usepackage{float}
\newcommand{\FigRef}[1]{Fig.~\ref{#1}}

\newcommand{\mytitle}{The dynamics of leadership and success in software development teams}

\begin{document}

\title{\mytitle}

\author{Lorenzo Betti}
\affiliation{%
 Department of Network and Data Science, Central European University, Vienna, Austria
}
\author{Luca Gallo}
\affiliation{%
 Department of Network and Data Science, Central European University, Vienna, Austria
}
\affiliation{
ANETI Lab, Corvinus Institute for Advanced Studies (CIAS), Corvinus University of Budapest, Budapest, Hungary
}
\author{Johannes Wachs}
\affiliation{
Institute for Data Analytics and Information Systems, Corvinus University of Budapest, Budapest, Hungary
}
\affiliation{
Institute of Economics, HUN-REN Centre for Economic and Regional Studies, Budapest, Hungary
}
\affiliation{
Complexity Science Hub, Vienna, Austria.
}
\author{Federico Battiston}
\email{battistonf@ceu.edu}
\affiliation{%
 Department of Network and Data Science, Central European University, Vienna, Austria
}

\begin{abstract}

From science to industry, teamwork plays a crucial role in knowledge production and innovation. 
Most studies consider teams as static groups of individuals, thereby failing to capture how the micro-dynamics of collaborative processes and organizational changes determine team success. 
Here, we leverage fine-grained temporal data on software development teams from three software ecosystems -- Rust, JavaScript, and Python -- to gain insights into the dynamics of online collaborative projects. 
Our analysis reveals an uneven workload distribution in teams, with stronger heterogeneity correlated with higher success, and the early emergence of a lead developer carrying out the majority of work. 
Moreover, we find that a sizeable fraction of projects experience a change of lead developer, with such a transition being more likely in projects led by inexperienced users. 
Finally, we show that leadership change is associated with faster success growth. 
Our work contributes to a deeper understanding of the link between team evolution and success in collaborative processes.

\end{abstract}

\maketitle

\section{Introduction}
\label{sec:intro}

The production of innovation and knowledge increasingly relies on collective efforts.
For example, teamwork is crucial in scientific research, where teams have demonstrated their effectiveness in fostering groundbreaking discoveries~\cite{wuchty2007increasing,guimera2005team, wu2019large}, or in industry, where teamwork is essential for the rapid development of innovative solutions~\cite{liu2015exploring}.
The effectiveness of teamwork depends on the continuous integration of specialized knowledge of individuals~\cite{menahem2016coordinating,grand2016dynamics}, and the ability to leverage constructive conflicts to generate novel insights~\cite{xiao2014reexploring,oneill2013examining}.
Extensive research has investigated characteristics of team members that are associated with performance~\cite{mariani2024collective}, from the effect of team size~\cite{wu2019large}, the interdependencies among team members~\cite{baumann2024network}, and their diversity in terms of  gender~\cite{yang2022gender,bear2011role,wallrich2024relationship,zeng2016differences,duch2012possible}, expertise~\cite{vaan2015game}, prior experiences~\cite{guimera2005team,pobiedina2013ranking}, and ethnicity~\cite{freeman2015collaborating,hoogendoorn2012ethnic}. 
In this way, the results of such collaborative efforts often go beyond the sum of individual contributions thanks to the emergence of synergies among team members~\cite{woolley2010evidence,ungar2012good}, conditional on the team's organization.

In a team, not all members are equal~\cite{arrow1993membership}. 
Successful teamwork often relies on the management of tensions by leaders~\cite{saeed2014leadership,vedres2020open}, who organize work through the division of complex tasks into sub-tasks among different team members~\cite{zaccaro2001team}. Even in self-organizing teams, lacking a predefined hierarchical structure, specific team members can emerge as leading figures who carry out a sizeable fraction of the work and are eventually responsible for the project advancement~\cite{omahony2007emergence,thapa2020evaluating,taggar1999leadership}. 
The organizational dynamic and evolving nature of teams impact team structure over time, as roles and responsibilities shift to adapt to new challenges~\cite{simon2013administrative,arrow2000small}. 
For instance, adjustments in the distribution of labor, the emergence and shift of leadership, and the turnover of team members all hinder coordination in teams, explaining why social and coordination skills are growing in value as teams get larger and teamwork more prevalent \cite{deming2017growing}.

Indeed, various conceptual frameworks have highlighted the need to explicitly consider team dynamics to properly understand how teams function and their success~\cite{arrow2000small,delice2019advancing, gorman2017understanding}.
Nevertheless, most of the insights we have consider teams as static entities, with little empirical research accounting for their temporal evolution. 
This is largely due to difficulties in accessing or collecting temporal data tracking and measuring team activities across time~\cite{humphrey2014team,santolini2023igem}.
In science, benefiting from the availability of large-scale curated publication data, a few studies have recently explored the concept of ``persistent teams''~\cite{chowdhary2024team} -- researchers who consistently collaborate together over time -- showing that ``fresh'' teams made of new collaborators produce more impactful research~\cite{zeng2021fresh}.
However, publication data only records the outcome of teamwork, lacking information about the process of the collaborative effort, such as the specific contributions and activities of the authors of each paper. 
An exception is the analysis of email communications within research teams, which can provide insights into individual contributions and team dynamics~\cite{duch2010quantifying}.

In organization theory, team activity is typically tracked through multi-period observations, where team members are surveyed at various intervals. 
Yet this approach suffers from inherent limitations of low temporal resolution because it would necessitate frequently surveying team members.
Such frequent surveys can disrupt teamwork and lead to survey fatigue, eventually compromising data quality and preventing the effective investigation of team dynamics at a microscopic scale~\cite{humphrey2014team}.
While fine-grained data about team processes can be collected through other methods like sensors that generate real-time data~\cite{kolbe2019laborious} or in controlled laboratory experiments~\cite{szabo2022anatomy}, these approaches typically cover only a limited number of teams and are difficult to scale.

Open-source software development offers an ideal opportunity to investigate collaborative dynamics at a large scale and with a fine-grained resolution \cite{scholtes2016aristotle,gote2021analysing,sornette2014how,zoller2020topology}.
In this context, software developers typically track changes to the software codebase using version control systems (e.g., Git) and store their software in online public repositories (e.g., GitHub).
For instance, by analyzing the contribution history of a project via its commit log (a record of every edit made to code), it is possible to obtain information about who made which modification to the software and when.
The analysis of this source of data has identified patterns where few developers often perform the majority of contributions in a project and most developers make few~\cite{goeminne2011evidence,yamashita2015revisiting,klug2016understanding}.
This tendency towards centralization, widespread across collaborative software development, has been employed to develop heuristics to identify the core members of teams~\cite{bock2023automatic}.
Changes in the activity of core developers, as well as changes in its composition, can offer novel insights about how teams function and the relationship to project success~\cite{zanetti2013rise}.
The dynamics of the main developers may vary significantly across projects.
For instance, Linus Torvalds, who created the Linux kernel in 1991, still dominates the development effort of the project~\cite{linuxcommits} and defines himself as a ``benevolent dictator of Planet Linux''~\cite{articlelinux}.
On the other hand, Pandas' creator Wes McKinney stopped actively working on Pandas after five years~\cite{articlepandas} and other developers have succeeded as the main developers of the project~\cite{pandascommits}. 
While both Linux and the Pandas library are examples of successful projects, their contrasting leadership dynamics highlight a gap in our understanding of the causes and consequences of the emergence of key team members, their potential turnover, and the effect of such changes on project success.

To advance the understanding of team dynamics in open-source software projects, we analyze teams' activity within three software ecosystems -- Rust, JavaScript, and Python -- along with their success metrics.
First, we examine the distribution of work as reflected by the distribution of commits among team members, study the activity patterns of the lead developers -- users responsible for the largest share of commits -- and correlate it with project success.
Then, we identify repositories where the lead developer changes during the project's lifespan, identifying profound redistribution of workload and a potential reorganization of the team. 
Finally, we investigate the association between the change of lead developer and the success growth of the project after the transition by comparing those repositories to similar ones that did not change the lead developer. 
Our findings, consistent across the three software ecosystems, demonstrate the interplay of team dynamics and performance in open-source software projects, suggesting that changes in team organization have implications for the success of the project.

\section{Results}
\label{sec:results}

We leverage three datasets tracking the activity of software developers working on GitHub repositories hosting packages from three popular programming languages: Rust, JavaScript, and Python.
For Rust, we rely on a well-curated dataset covering over \num{40000} developers working on more than \num{6000} repositories between 2014 and 2022, including their success time series~\cite{schueller2022evolving,schueller2024modeling}.
The JavaScript and Python datasets are about four and two times bigger, respectively, with activity from 2010 to 2018~\cite{marat2018ecosystem, marat2018ecosystemData}. We supplement these two datasets by collecting success time series data (see Section~\ref{app:replication_analysis} in Supplementary Material).

In the main text, we present results focusing on the Rust ecosystem because it offers the greatest level of detail and coverage of the three systems.
Indeed, as Rust was released in 2015, we have access to comprehensive temporal data on project development and GitHub-specific metrics such as stars. Moreover, the Rust package manager records download data since release.
Javascript and Python, on the other hand, were created in the 1990s, well before the creation of GitHub, which limits our ability to collect comprehensive data on these languages. Similarly, data on downloads for Python and Javascript libraries are available only for limited time periods.
Despite these limitations, we can still run our analyses in these much larger and longer-running systems and our substantive findings on team dynamics and project success are consistent across all three languages.
We provide detailed information for JavaScript and Python in Section~\ref{app:replication_analysis} of Supplementary Material.

\subsection{Emergence of a lead developer}
\begin{figure*}[t]
    \centering
    \includegraphics[width=\textwidth]{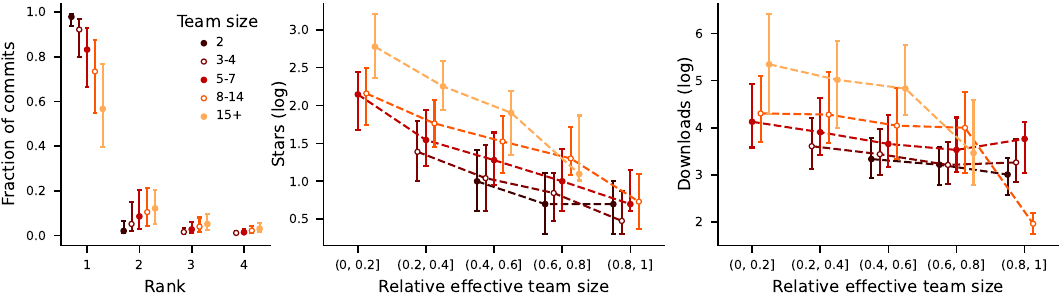}
    \caption{\textbf{Workload distribution within teams and relationship with success.} 
    (\textbf{a}) Median fraction of commits authored by the $r$-th most active developer of a repository stratified by team size. 
    The most active developer makes more than half of the total number of commits while other developers contribute substantially less, regardless of the size of the team.
    (\textbf{b-c}) Median number of stars (\textbf{b}) and downloads (\textbf{c}) as a function of the relative effective team size stratified by team size. 
    The more heterogeneous the workload distribution in the team, the higher the success. 
    The Spearman's rank test returns $p < 0.001$ for all team sizes. 
    The number of stars and downloads are incremented by one unit.
    Error bars range from the 25th to the 75th percentile of the distributions.
    }
    \label{fig:heterogeneity_work_distrib}
\end{figure*}

Software developers track their changes to the software codebase through commits, whose distribution across team members can measure their work contributions~\cite{klug2016understanding} and be informative about their roles in a project~\cite{bock2023automatic}.
After having verified that most of the commits refer to actual coding activities (see \FigRef{Sfig:distrib_commit_types}), 
we begin by analyzing the distribution of commits among team members to characterize the distribution of work in software development teams.
We define a team as the set of developers who make at least one commit to a repository.
By ranking the developers of a repository by their total number of commits, we measure the fraction of commits authored by developers as a function of their rank.
\FigRef{fig:heterogeneity_work_distrib}a shows that the most active developer ($\text{rank}=1$) authors more than half of the total number of commits.
In contrast, the second most active developer typically accounts for only around 10-20\% of the total, while the rest is done by the other team members.
This observation highlights the presence of a ``lead developer'' who carries the majority of the workload in a repository, alongside other developers contributing to a lesser extent.
Those properties are consistent across teams of different sizes as shown in \FigRef{fig:heterogeneity_work_distrib}a (see \FigRef{Sfig:distribution_team_size_and_releff_team_size}a for the distribution of repositories across different team sizes), through the lifetime of repositories (see \FigRef{Sfig:heterogeneity_work_distrib_year}), and cannot be explained by random activity of developers (see \FigRef{Sfig:null_heterogeneity_workload_distribution}).
This persistent nature of the workload distribution suggests a possible advantage from such a centralization.

One potential aspect where this advantage may manifest is projects' success.
To test for a relationship between heterogeneous workload distribution and success, we first quantify
the heterogeneity of the workload distribution using the relative effective team size~\cite{klug2016understanding}.
This metric is defined as:

\vspace{-0.5cm}
\begin{equation*}
    \text{rel. eff. team size} = 2^{H}/N
\end{equation*}
where $H$ is the binary entropy of the distribution of commits among team members and $N$ is the size of the team. This measure ranges from $1/N$ (i.e., one single developer makes all commits) to $1$ (i.e., the workload is evenly distributed among all members). 
Consequently, a smaller relative effective team size indicates a more uneven distribution of commits among team members.
Then, we employ the number of stars and downloads as metrics of repositories' success.
Such metrics track two different dimensions of success: stars can be considered as a proxy for repositories' popularity, similar to likes in social media~\cite{borges2018what}, whereas the number of downloads reflects factors such as utility, necessity, and perceived quality.
Our analysis reveals an inverse relationship between repositories' success and their relative effective team size, as displayed in \FigRef{fig:heterogeneity_work_distrib}b for the number of stars and \FigRef{fig:heterogeneity_work_distrib}c for the number of downloads (see \FigRef{Sfig:distribution_team_size_and_releff_team_size}b for the distribution of the number of repositories for different relative effective team sizes).
The relationship is supported by Spearman's correlation test, yielding a correlation coefficient of at least $\rho=-0.28$ for stars and $\rho=-0.11$ for downloads across all team sizes ($p<0.001$; see Table~\ref{tab:corr_success_rets} for details). 
Note that we grouped repositories according to their team size since this is already strongly correlated with success (see \FigRef{Sfig:corr_team_size_success}).
To check if the age of repositories affects the results of the correlation, we repeated the analysis considering the relative effective team size and success at different moments of repositories' lifetime, finding consistent results (see \FigRef{Sfig:heterogeneity_work_distrib_year} and Table~\ref{tab:corr_success_rets}).

Our observation aligns with prior research on software development teams~\cite{yamashita2015revisiting}, confirming that even in a relatively new programming language such as Rust, the most successful teams of software developers have an uneven workload distribution~\cite{klug2016understanding}.
Same findings hold for JavaScript and Python as shown in \FigRef{fig:heterogeneity_work_distrib_js_py}.
Furthermore, as already found in~\cite{goeminne2011evidence} where three large open-source projects were studied, our results show that the workload distribution becomes heterogeneous already within the first year of activity. 
Differently from previous studies, in the following we provide a characterization of the behaviors of the lead developers of each repository.  
Beyond static analyses, we focus in particular on team dynamics, revealing changes in workload distribution and how this impacts a project's success.

\subsection{Characterization of lead developers' activity}

Lead developers are not just prolific team members, but they fulfill specific management and coordination roles within their projects.
The first aspect that underscores a distinguished role is whether they have direct write privileges, granting them the authority to integrate changes into the repository.
We found that all identified lead developers do indeed have this right (see Section~\ref{app:write_access} in Supplementary Material).
This privilege extends beyond integrating their own work, but it also confers the authority to review and decide whether others’ contributions are worth being accepted.
By searching for commits related to merging pull requests -- the principal mechanism of collaboration in GitHub -- we revealed that most lead developers are heavily involved in such merging activities (83\% of them; see Section~\ref{app:mergers} in Supplementary Material), suggesting that they review and decide which contributions to accept.
Finally, coordinating team efforts is critical to managing a project effectively. 
We therefore assessed lead developers’ roles in coordination by analyzing their centrality in the communication networks built from discussions under issues and pull requests, finding that they frequently occupy the central positions (in 82\% of the cases they are among the top-three; see Section~\ref{app:coordination_leads} in Supplementary Material). 
Taken together, these observations reveal that lead developers fulfill core leadership tasks, from integrating and reviewing external contributions to coordinating team efforts.

\begin{figure*}[t]
    \includegraphics[]{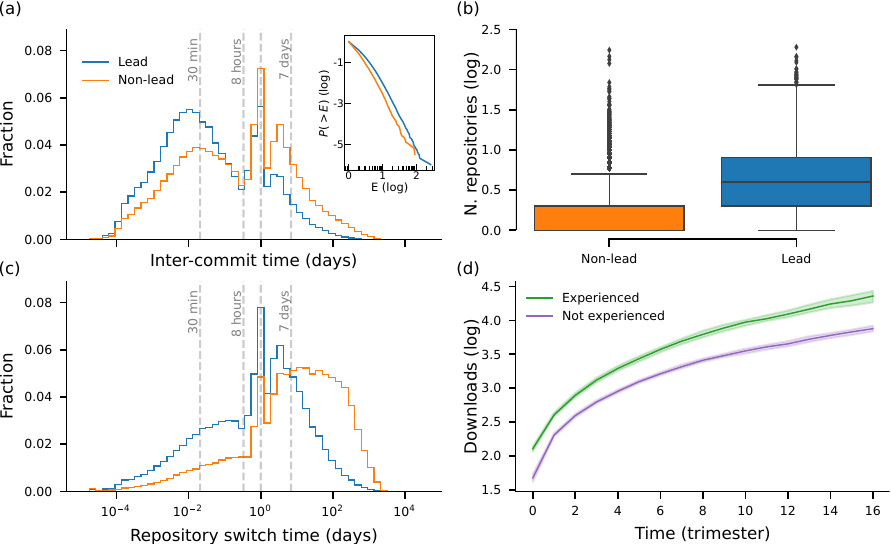}
    \caption{\textbf{Characterization of lead developers' activity.} 
    (\textbf{a}) Distribution of the inter-commit times and cumulative distribution of the number of commits close in time (inset) for lead and non-lead developers. 
    Lead developers exhibit higher frequency of commits and longer streaks of consecutive commits.
    (\textbf{b}) Distribution of the number of repositories in which lead and non-lead developers are active. 
    Lead developers are involved in a larger number of repositories.
    Box plots indicate median (middle line), 25th, 75th percentile (box) and 5th and 95th percentile (whiskers) as well as outliers (single points).
    (\textbf{c}) Distribution of the repository switch time of lead and non-lead developers.
    Lead developers tend to switch from one project to another on a daily to weekly basis.
    (\textbf{d}) Number of downloads across repositories' lifetime stratified by lead developers' experience (median and its 95\% confidence interval). 
    Repositories led by experienced developers are downloaded more compared to those led by inexperienced ones.
    Time is binned into trimesters.
    }
    \label{fig:leaders_activity}
\end{figure*}

We now shift our focus to how the coding activities of leaders differ from those of other developers.
We compare the distribution of their inter-commit time, defined as the time elapsed between two consecutive commits authored by the same user (regardless of the repository in which the commit is made).
\FigRef{fig:leaders_activity}a displays the distributions for lead and non-lead developers separately, showing distinct characteristics for those two sets of users.
In particular, lead developers exhibit a higher frequency of commits.
Indeed, their inter-commit time distribution is left-skewed compared to that of non-lead developers, displaying a prominent peak around 30 minutes (the peak in non-lead developers' inter-commit time ranges between one day and one week).
This difference turns into lead developers being more likely to have longer streaks of consecutive commits, as depicted in the inset of \FigRef{fig:leaders_activity}a.
The inset shows the probability distribution of making more than $E$ consecutive commits whose inter-commit time is smaller than $\Delta t = 30$ minutes (the observation is robust across other values of $\Delta t$ as can be seen in \FigRef{Sfig:ccdf_train_events}).
Lead developers are indeed three times more likely to make streaks of at least 10 commits ($P\left(\geq 10\right) = 0.0097$) compared to non-lead developers ($P\left(\geq 10\right) = 0.0030$).

Another aspect of the activity pattern of developers is the number of repositories they contribute to.
We first show in \FigRef{fig:leaders_activity}b the distribution of the number of repositories in which lead and non-lead developers author at least one commit (see \FigRef{Sfig:developers_experience_and_success} for the aggregate distribution).
In addition to being the most active developers in their repositories, lead developers contribute to a median of 4 repositories (interquartile range IQR 2-8) while non-lead developers are active in only a median of 1 repository (IQR 1-2).
This difference is deemed significant according to the Mann–Whitney \textit{U} test ($U=\num{96197128}$, $n_1=\num{3627}$ and $n_2=\num{32998}$, $p<0.0001$).
To gauge insights on the extent to which lead developers work on multiple projects simultaneously, we introduce the repository switch time of developers.
This is defined as the time elapsed between a developer's first commit on a repository and their first commit on a different one, namely the time elapsed between a developer initiating work on one repository and transitioning to another.
\FigRef{fig:leaders_activity}c shows the distribution of repository switch time for lead and non-lead developers separately.
The difference with the inter-commit time distribution in \FigRef{fig:leaders_activity}a suggests that the trains of consecutive commits shortly separated in time (i.e., inter-commit time of 30 minutes) are mostly done on the same repository, as lead developers switch from one repository to another on a daily to weekly basis.
The rapid decay of the repository switch time after one week, in addition to the number of repositories in which lead developers are active, implies that they work concurrently on multiple projects and focus their efforts within the Rust ecosystem.
Non-lead developers rather move between repositories over longer periods. This may be a consequence of non-lead developers being active in fewer projects and possibly being less involved in the Rust ecosystem.

Lead developers often contribute to multiple repositories, and their experience as core members in previous projects might prove beneficial for new endeavours. 
For this reason, we ask: Does an experienced lead developer impact the success of a repository?
We define the lead developer of a repository $i$ as experienced if they have been the lead developer of other repositories before starting to contribute to repository $i$.
We identify \num{2744} (43\%) repositories led by experienced lead developers and show in \FigRef{fig:leaders_activity}d the median number of downloads these repositories receive across their lifetime, compared to repositories whose lead developer never led a repository before.
We observe that repositories led by developers with previous experience receive more downloads over their lifetime. 
Since downloads reflect factors such as utility, necessity, and perceived quality, we speculate that developers who led repositories in the past may leverage their previous experience in Rust projects to develop software that is more useful and widely needed.
By contrast, it is worth noticing that lead developers' previous experience does not affect the number of stars, indicating a lack of advantage in terms of popularity (see \FigRef{Sfig:success_vs_time}).
We verified that the observations are not influenced by the inclusion of repositories with varying ages, as this could introduce bias due to the advantage older repositories may have in accumulating stars and downloads (see \FigRef{Sfig:success_vs_time}).

All these observations are mirrored in the other two datasets.
Specifically, we show the distribution of inter-commit times in \FigRef{fig:leaders_activity_js_py}~(a,e), the distribution of the number of repositories where developers are active in \FigRef{fig:leaders_activity_js_py}~(b,f), the distribution of repository switch time in \FigRef{fig:leaders_activity_js_py}~(c,g), and the success across time depending on the experience of the lead developer in \FigRef{fig:leaders_activity_js_py}~(d,h).

\subsection{Lead developers of repositories can change}

\begin{figure*}[t]
    \includegraphics[width=\textwidth]{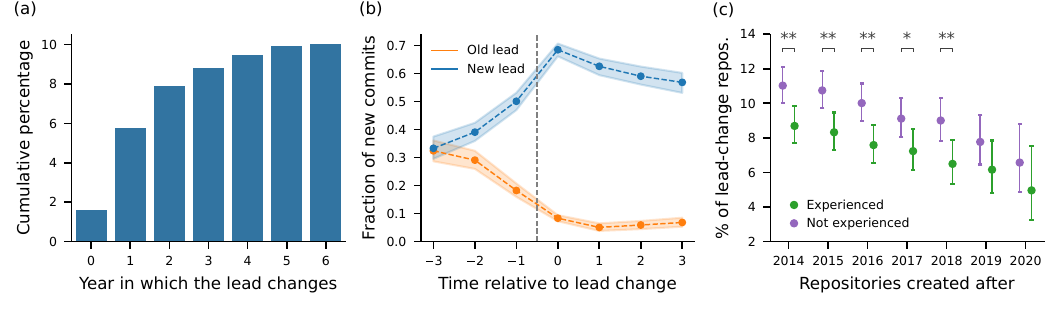}
    \caption{\textbf{Lead developers can change across the lifetime of repositories.} 
    (\textbf{a}) Cumulative percentage of repositories undergoing a lead developer change as a function of the number of years since their creation.
    Around 10\% of repositories change their lead developer throughout their lifetime, with the majority occurring within the second and third year of activity.
    (\textbf{b}) Fraction of new commits authored by the old and new lead developer before and after the lead developer transition (mean and its 95\% confidence interval).
    After the transition (vertical dashed line), contributions from the old lead developer diminish rapidly.
    (\textbf{c}) Percentage of lead-change repositories stratified by the previous experience of the old lead developer.
    Each point refers to repositories created at a specific year or later.
    Repositories led by inexperienced lead developers exhibit a significantly higher likelihood to change their lead developer compared to those led by experienced ones, according to Fisher's exact test.
    Significance levels are denoted as follows:
    * for $p < 0.05$, ** for $p < 0.01$, and *** for $p < 0.001$.
    Error bars refer to 95\% confidence intervals of the estimated percentages (Wilson score interval).}
    \label{fig:leader_changes}
\end{figure*}

Lead developers emerge early in a project and display distinct activity patterns.
Does the same individual consistently maintain such a role, or does this figure change in time?
To identify such changes, we first aggregate developer activities (i.e., cumulative number of commits) and the success of repositories (i.e., cumulative number of stars and downloads) by trimesters.
Then, we consider the lead developer of a repository at trimester $t$ as the one who has made the largest share of commits up to that time (we excluded 206 repositories from the subsequent analyses to ensure the inclusion of repositories displaying meaningful changes; see details in the Methods section).
We find that \num{618} repositories (10\%) undergo a change of lead developer, with the majority of such changes occurring between the second and third year of a repository's lifetime, as shown in \FigRef{fig:leader_changes}a.
Most repositories undergoing a change in lead developer experience only one transition (92\%).
Therefore, we focus our subsequent analyses on the first and second lead developers, hereafter referred to as the old and new lead developers, respectively.

To further investigate how the transition from the old to the new lead developer occurs, the first aspect we examine is the dynamics of the transition.
We show in \FigRef{fig:leader_changes}b the relative number of new commits authored by the old and new lead developer around the time when the transition occurs (i.e., the time when the lead developer changes), averaged across repositories (repositories with no activity in a trimester are omitted).
The result indicates a rapid transition with a drop in the activity of the former lead developer.
One year before the transition, old and new lead developers make a comparable amount of new contributions on average, i.e., 30\% of the total number of new commits.
However, while the activity of old lead developers sharply declines, dropping to levels below 10\%, the relative contribution of new leads largely increases, ultimately making a heavy portion of new commits (around 65\%) in the subsequent year after the change.
In particular, in 297 repositories, corresponding to 48\% of projects that undergo a change, the old lead developer ceases to contribute entirely through commits. 
This stop in commit activity does not necessarily indicate a complete departure from the project, as contributors can remain involved in other activities within projects. 
Indeed, we observe that in 48 of such repositories (16\%), old lead developers remain active by engaging in other tasks such as addressing issues and reviewing pull requests.
This transition of core team members to administrative roles has been previously observed in open-source software development as a response to an abrupt increase of external attention~\cite{maldeniya2020herding}, and reflects in new lead developers moving to more central positions in the communication network of the project (see Section~\ref{app:coordination_leads} in Supplementary Material).

Next, we ask if previous experience can explain changes to the role of the lead developer.
To account for the temporal dimension, we consider the lead developer of a repository $i$ as experienced if they have been the lead developer of other repositories before becoming the lead developer of repository $i$.
We find that 11\% of repositories initiated by inexperienced lead developers undergo a transition of lead developer (388 out of 3519), while this happens to 9\% of those initiated by experienced lead developers (230 out of 2646).
Although this difference looks small, it corresponds to a 30\% increase in the odds of changing the lead developer when the initial lead lacks previous experience and it is deemed significant according to Fisher's exact test (odds ratio at 1.30, $p\text{-value} = 0.003$).
We checked the robustness of the result against the year in which repositories were initiated by conducting additional tests restricted to repositories initiated after specific years.
The results, shown in \FigRef{fig:leader_changes}c, confirm that the association remains significant except for the last two years, possibly because of the reduction of the sample size ($N=2339$ for 2019 and $N=1027$ for 2020).
In short, the experience of the initial lead developer is associated with a lower likelihood of change.

Finally, we investigate if lead developers changes may be explained by the success of the project before the change.
For instance, existing literature on startups suggests a U-shape relationship between founder departure and growth rate, indicating that founders of startups experiencing either slow or rapid growth are more likely to depart than those of startups with intermediate growth rates~\cite{boeker2002entrepreneurial}.
Interestingly, we find no evidence supporting an association between previous success and the likelihood of changing the lead developer in the future within the Rust ecosystem (see \FigRef{Sfig:leader_change_vs_prev_success} for details).

These results replicate almost entirely in the JavaScript and Python datasets.
We provide the cumulative fraction of repositories undergoing a lead developer change in \FigRef{fig:leader_changes_js_py}~(a,d), the activity dynamics close to the transition point in \FigRef{fig:leader_changes_js_py}~(b,e), and the likelihood of lead developer change depending on the experience of the lead developer in \FigRef{fig:leader_changes_js_py}~(c,f).
The main difference is the weaker yet significant association between lead developer change and experience in the JavaScript dataset (odds ratio at 1.11, $p=0.02$).

\subsection{Repositories that change the lead developer perform better after the change}

With the previous analyses, we have shown that a sizeable fraction of repositories undergo a change of their lead developers, describing the dynamics and factors associated with this turnover.
How does such a turnover relate to the repositories' future success?

To answer this question, we employ a matching approach to compare the success trend of those repositories (named ``lead-change repositories'') against the success of similar repositories whose lead developer did not change (named ``lead-remain repositories'').
Specifically, we design a stringed matching procedure to identify pairs of lead-change and lead-remain repositories that are similar in terms of team composition and success prior to the change of the lead developer (see Methods section for details).
To ensure a reasonable fit, we restrict the analysis to a subset of 151 lead-change repositories (24\%) displaying meaningful activity and success one year before the change of lead developer (more than 50 commits, 10 stars, and 100 downloads).
We then monitor the difference in the success growth at time $t$ ($\Delta_t$) associated to the lead developer change as

$$ \Delta_t = Y_t - \Tilde{Y_t} $$
where $Y_t = S_t / S_{t_0}$ is the success of the lead-change repository (i.e., either number of stars or downloads) relative to the last pre-treatment period ($t_0 = -1$) and $\Tilde{Y_t}$ is that of the matched repository.
The value of $\Delta_t$ should be close to zero for $t \leq t_0$, indicating that the lead-change repository and its match exhibit similar success trajectories before the change.
After $t_0$, $\Delta_t$ quantifies how the success growth varies in relation to the change of lead developer. 
We consider $\Delta_t$ for $t$ ranging from $\left[-4, 4\right]$, namely, one year before and after the time in which the change occurs ($t=0$).

\begin{figure*}[t]
    \includegraphics[]{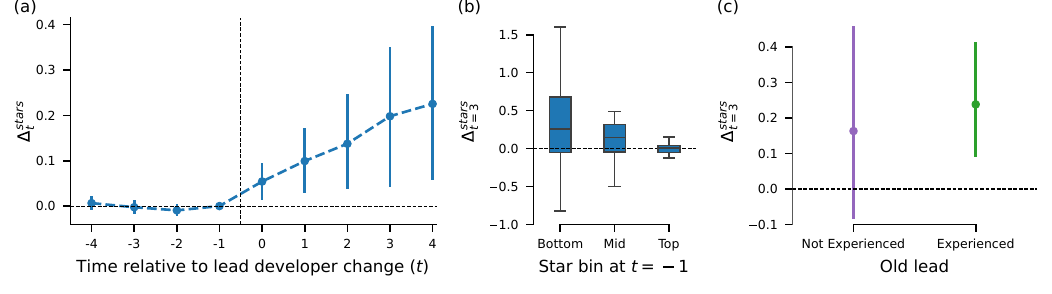}
    \caption{\textbf{Lead developer changes are associated with faster success growth.} 
    (\textbf{a}) Average effect of lead developer change $\Delta_t$ for stars. 
    Repositories' success grows faster compared to similar repositories that did not undergo such a change. 
    (\textbf{b-c}) Success growth $\Delta_t$ for stars stratified by (\textbf{b}) the success before the change of lead developer and (\textbf{c}) the experience of the old lead developer.
    Worst-performing repositories exhibit a large positive effect following the change, whereas top-performing ones are minimally affected.
    Repositories started by an experienced lead developer benefit more from the change than those initiated by an inexperienced one.
    Error bars in (\textbf{a}) and (\textbf{c}) correspond to 95\% confidence intervals of the estimated quantities.
    In (\textbf{b}) box plots indicate median (middle line), 25th, 75th percentile (box) and 5th and 95th percentile (whiskers) as well as outliers (single points).
    }
    \label{fig:leader_changes_correlation_success}
\end{figure*}

We find that the change of lead developer is positively related to a stronger growth in success.
\FigRef{fig:leader_changes_correlation_success}a shows the success difference $\Delta_t$ for stars averaged across the 135 repositories with a suitable match (89\% of the lead-change repositories). 
Notably, the difference in success growth is already positive during the first trimester in which the new lead developer takes over ($\Delta_{t=0} = 0.05$ on average, 95\% CI: 0.02 - 0.09) and keeps increasing during the whole year after the change (one year later, $\Delta_{t=4} = 0.23$ on average, 95\% CI: 0.06 - 0.40).
Same results hold for downloads (see \FigRef{Sfig:effect_leader_change_mean_median}), thus suggesting that the change of lead developer is associated with faster success growth with respect to both popularity and perceived utility and quality of the software.
The values of $\Delta_t$ before the change ($t<0$) are close to zero, meaning that the success trends between the lead-change and their matched repositories are close on average before the transition.
This underlines the good quality of the matching procedure.
To check the robustness of the result, we also consider the median of the success difference $\Delta_t$, showing that the observed positive association is not due to the skewness of the distribution of $\Delta_t$ (see \FigRef{Sfig:effect_leader_change_mean_median}).
Moreover, we verified that the observed association is not confounded by the heterogeneity in the skills and abilities of lead developers by performing a fixed effect regression, which yielded consistent results (see Supplementary Table~\ref{Sfig:panel_regression_leader_fe}).
This finding is consistent for JavaScript and Python in both direction and magnitude, as shown in \FigRef{fig:leader_changes_correlation_success_js_py}~(a,d).

Finally, we investigate how success before the transition and lead developer experience affect success growth.
As depicted in \FigRef{fig:leader_changes_correlation_success}b, we find that lowly successful repositories (i.e., those in the bottom 30\% of the success distribution) have larger $\Delta_t$ in terms of number of stars when compared to average (i.e., those between the 50th and the 80th percentile) and top (i.e., top 10\%) repositories (Kendall's $\tau_b=-0.18$, $p = 0.03$).
This indicates that repositories gaining relatively more visibility after the change of lead developers are the least popular.
We found no significant differences in the case of downloads (Kendall's $\tau_b=-0.01$, $p = 0.92$; see \FigRef{Sfig:het_effect_previous_success}).
Regarding the previous experience, we observe in \FigRef{fig:leader_changes_correlation_success}c that both repositories initiated by an experienced or not experienced lead developer have a significant growth in terms of stars (Not experienced: Wilcoxon's $T=1682$, $n=72$, $p=0.02$, one sided; Experienced: Wilcoxon's $T=1488$, $n=63$, $p=0.0004$ one sided) (see \FigRef{Sfig:het_effect_leader_experience} for a discussion of the skewness of $\Delta_t$).
Finally, in \FigRef{Sfig:het_effect_both_leader_experience}, we consider the combined effect of both the old and the new lead developer's experience on success growth. 
We note that these observations are not in line with what found for JavaScript and Python.
Specifically, we observe no difference in $\Delta_t$ for different levels of success before the transition (see \FigRef{fig:leader_changes_correlation_success_js_py}~(b,e)) and mixed results for what concerns the differential effect of lead developers' experience on $\Delta_t$ (see \FigRef{fig:leader_changes_correlation_success_js_py}~(c,f)). 
These discrepancies may be attributed to intrinsic differences of the software ecosystem under study.

\section{Discussion}
\label{sec:discussion}

In this study, we used fine-grained data about software development teams from three popular programming languages to characterize the temporal dynamics of online collaborative projects.
By tracking the activity of the most active contributor, we unveiled the emergence of a lead developer who exhibits a distinctive pattern of activity compared to non-lead developers, with lead developers emerging early in a project lifetime. 
Moreover, we show how an uneven distribution of workload among team members is positively associated with a repository's success.
We identified a sizeable fraction of projects that undergo a change of their lead developer and revealed an association between such transitions and faster success growth. 

Our analysis of open-source projects reports that most of the work is carried out by one or a few developers and that this correlates with higher success.
This evidence is well-documented in the literature and spans across different programming languages~\cite{mockus2002case,goeminne2011evidence,chelkowski2016inequalities,yamashita2015revisiting,klug2016understanding}.
One possible explanation for such a correlation is the increase in efficiency due to the concentration of workload among few developers, which is likely to reduce the cost of coordination~\cite{ye2007reducing,klug2016understanding}.
At the same time, this distribution of work may result in a concentration of knowledge about the functioning of the software around a few developers, thus posing the project at risk should the main developers leave.
Drawing on a well-known concept in the software development literature, projects have a small ``truck factor'', meaning that the number of key developers who would need to be incapacitated, i.e., hit by a truck, to prevent further development of the project is small~\cite{avellino2016novel,pfeiffer2021identifying,williams2003pair,bowler2005truck,ferreira2019algorithms}.
The tension between efficiency and project sustainability highlights a ``high risk, high reward'' strategy in open-source software development.

Moving beyond the open-source software development literature, whether a hierarchical or flatter organization fosters better team outcomes is a long-standing debate, summarized by the tension between the functionalist  and the conflict perspectives. 
The former posits that team outcomes benefit from a hierarchical structure because it facilitates coordination, whereas the latter predicts a negative effect because of tensions arising from non-aligned goals across the levels of the hierarchy~\cite{greer2018why}.
Many empirical and theoretical efforts further demonstrate that such a relationship may depend on task type, task complexity, and additional mediators~\cite{brackbill2020impact,almaatouq2021task,lazer2007network,barkoczi2016social,centola2022network,greer2018why}.
These cross-cutting observations underscore the lack of an overarching understanding of this relationship and motivate further comparative studies to examine the interplay of domain specifics, team dynamics, and task type on team outcomes. 
Our findings add to this wider conversation by focusing on open-source development, where uneven team structures appear widespread and are linked to higher success.
Notably, recent analyses of top-starred GitHub repositories indicate that highly successful projects often exhibit pronounced hierarchical collaboration structures, suggesting that such patterns of division of labor can indeed thrive in large, complex projects~\cite{palazzi2019online}.  
Such analyses focus only on a few, large projects, and are difficult to scale~\cite{gote2022big}. 
While our broader, commit-level approach enables an extensive coverage of projects, we acknowledge that it does not allow to detect these nuanced hierarchies or reveal the specialized or modular collaboration patterns that may emerge in the largest teams.

More broadly, our study contributes to the ongoing discourse within the field of team science, particularly by addressing the unresolved question of how changes in team composition affect teamwork~\cite{humphrey2014team}. 
Although the literature presents mixed findings, where team changes can be either beneficial~\cite{gorman2010team,choi2005old} or detrimental~\cite{ramos2012chaotic,lewis2007group}, there is consensus that changes involving core team members deeply affect team functionality~\cite{arrow1993membership}.
Our results are not only in line with these general predictions, but lean towards a positive effect of such changes on team outcomes.
In addition, the stronger effect observed for experienced lead developers of Rust projects leaves room to speculate about the importance of experienced developers in setting the stage for successful projects regardless of potential changes in such a fundamental role.
This observation raises further questions about the characteristics of the old and new lead developers that may influence such a relationship.
Drawing on previous studies, the change of a team member is more beneficial the higher the relative skills of the team members involved in the turnover~\cite{summers2012team}.
Further research is needed to identify factors that make the turnover more beneficial for team outcomes.
Additionally, our study mainly focus on the figure of the lead developer but studying the turnover and eventual shifts in workload distribution among the other team members remains an interesting direction for future research.

While our findings relate the change of the lead developer with a significantly higher success growth, the mechanism behind this association remains unclear. 
Indeed, this relationship highlights a deep connection between the dynamics of teams and their performance, which leaves room for speculation on whether success drives leadership change or vice versa.
For instance, as a result of a rapid increase of attention, projects can attract new contributors, and core developers can shift to organizational roles~\cite{maldeniya2020herding}. 
This suggests that an increase in success brings more developers to join the team, thus increasing the probability of changing the lead developer.
Since we consider three popular programming languages, a similar mechanism may be in place,
which likely increases the number of new contributors to projects and may require teams to adapt and undergo structural changes~\cite{galesic2023beyond}.
Conversely, leadership turnover can unleash creative forces previously bound up in an organization or team, which drive the increase in the project's success~\cite{tzabbar2014can}. 
At the same time, in our context we saw that new leaders tend to ramp up their activity in a project before taking over, suggesting that there may be value in a balance of previous collaborative ties and new connections in an evolving team \cite{vedres2020open}. 
In short, further research is needed to shed light on the causality between lead developer change and success.

Even though we analyzed three programming languages, we chose to keep their analyses separate and focus on Rust in the main text.
Besides the superior quality of the Rust dataset compared to the JavaScript and Python ones, we believe that our analyses benefit from considering the whole development history of one single programming language at a time, providing a controlled environment to characterize team dynamics in open-source software development.
Indeed, considering multiple programming languages together may affect the results because coding practices differ depending on the programming language.
For instance, it has been shown that different programming languages have varying levels of productivity and require unequal effort to write the same code~\cite{lavazza2016empirical,lutz2000empirical}.
In addition, we cannot have access to complete success time series for relatively old programming languages such as JavaScript and Python.
Indeed, the success metrics we considered (i.e., GitHub's stars and downloads) are platform-specific and may not be available for programming languages that are older than software-development platforms, as we may not have data on their repositories' entire development history.
Keeping these challenges and limitations in mind, we have been able to compare such ecosystems obtaining similar results (see Section~\ref{app:replication_analysis} of Supplementary Material).
The generalizability of our findings to three among the largest and most used programming languages indicates that our results may apply to OSS development in general.

Beyond open-source software development, our work provides a fresh perspective on team evolution, leadership dynamics and their relationship to project success, contributing to a deeper understanding of the successful dynamics of collaborative processes.

\section{Methods}
\label{sec:methods}

In this section, we focus again on the Rust dataset for the sake of consistency with the main text.
All the methods were applied consistently to the JavaScript and Python datasets.
The major differences regard the construction of the JavaScript and Python datasets, which are detailed in Section~\ref{app:replication_analysis_data_collect} of Supplementary Material.

\subsection{Data and selection of repositories}
We used data sourced from~\cite{schueller2022evolving}, consisting of a curated dataset containing the activities of developers across \num{39671} repositories hosting Rust packages on various online platforms (e.g., GitHub, GitLab).
Developers' activities are tracked through commits, providing a comprehensive record of changes to the project's codebase, in addition to other activities pertaining more to project management tasks (e.g., pull requests, Q\&As).
Developers' user names are disambiguated, and flags identifying bot accounts are provided.
The dataset includes platform-specific features that can be used as a proxy for repositories' success over time.
We chose to use the number of stars, which can be considered the most reliable measure of popularity~\cite{borges2018what}, and number of downloads, which reflects factors such as utility, necessity, and perceived quality.
Stars and downloads provide a multifaceted perspective on the success of repositories.
Since stars are only available for repositories stored on GitHub, we discarded those hosted in other platforms (6\% of the projects in the datasets).

Inspired by~\cite{gote2022big}, we filtered repositories to ensure the inclusion of repositories suitable to study collaborative software development.
After discarding the activity of bots, we selected the repositories satisfying the following conditions: (1) first commit with no deletions, (2) total number of lines of code positive across the whole lifetime, (3) first commit in 2014 or later, (4) at least 100 lines of code in total, (5) lifetime of at least one year, (6) at least one commit per month on average, (7) at least one package associated to the repository (since a repository can host more than one package~\cite{schueller2022evolving}).
Conditions (1) and (2) make us more confident that we study repositories for which their whole history is tracked.
Indeed, no repository can be initiated by deleting any line, nor can it have a negative number of lines at any point in its lifetime.
Conditions (4-7) select repositories hosting software developed over time and likely discard very small projects.
After discarding repositories developed by one single developer, and repositories displaying activity on less than four trimesters, we ended up with a total of \num{6165} repositories.

\subsection{Detecting lead developer changes}
The lead developer of a repository is defined as the developer with the highest number of commits.
However, different developers may be leading the repository at different points in time.
To identify changes in lead developers over time, we counted the cumulative number of commits authored by team members at each time period (i.e., trimesters).
Then, we defined the lead developer of the repository at time $t$ as the team member that made the largest share of commits up to that time.
To avoid the inclusion of spurious changes, such as those in the initial phases of the project when the activity is relatively low, we excluded the repositories that have changed the lead developer within the first three trimesters of activity.
Additionally, cases where a lead developer change is followed by the former lead developer re-assuming their role were excluded as well (total repositories discarded $N_\mathrm{spurious}=206$). 
This criterion ensures that our analyses rely on repositories in which a meaningful change happened, with the initial lead developer maintaining their position for a significant duration before being succeeded by a new lead developer.

We describe in Section~\ref{app:alternative_lead_detection} of Supplementary Material a more restrictive definition of lead developer that introduces a statistical test to determine whether the most active developer has a commit count that is significantly larger than the one of the second lead developers.
Our findings result robust to this alternative definition.

\subsection{Matching procedure}
To investigate the performance of repositories after the change of their lead developer (``lead-change repositories''), we compared their success trajectory with that of similar ones whose lead developer did not change (``lead-remain repositories").
Specifically, we implemented a matching procedure that, for each lead-change repository, identifies a set of lead-remain repositories that are similar to the lead-change repository in terms of temporal patterns of activity, team composition and success before the change of lead developer happened.
Then, we selected the matched repository among those candidates as the one with the most similar success trajectory during the year preceding the change of lead developer.

The details of the matching are as follows.
Initially, for each lead-change repository, we identified lead-remain repositories whose first commit date differs by at most six months from that of the lead-change repository.
In addition, we required the lead-remain repositories to have lifetime as long as the age at which the lead-change repository changed the lead developer.
This ensures that the lead-change repository and its candidates developed within a comparable timeframe.
Such a requirement contributes to controlling for the average status of Rust's ecosystem, thus avoiding potential biases due to the rapid growth of the programming language.
In this setting, $t=0$ designates the time in which the new lead developer takes over. 
Then, we refined our set of candidates to select repositories that closely matched the lead-change repository in terms of team composition and prior success for $t < 0$.
The selected set of candidates met the following criteria: (1) similar team size at $t=-1$ (see \FigRef{Sfig:similar_team_sizes_match}), (2) absolute difference in relative effective team size, averaged across $t\in \left[-4, -1 \right]$, smaller than 0.20, and (3) absolute relative difference of (log) number of stars and downloads at $t=-1$ smaller than 0.50.

As the last step, we selected the candidate that most closely matched in terms of success growth before the change.
To do that, we first defined the success growth as $Y_t = S_t / S_{t_0}$, where $t_0=-1$ refers to the last time period before the change of the lead developer and $S_t$ is either the number of stars or downloads at time $t$.
In other words, we considered the success growth relative to the success of the repository at the last trimester before the lead developer change. 
We then chose the matched repository as the one that exhibited the smallest maximum relative difference in success growth during the period $t \in \left[-4, -1\right]$, considering both stars and downloads.
If the maximum difference is larger than 0.50, we discarded the lead-change repository due to the lack of a sufficiently similar repository among the lead-remain ones.
Our results are robust against the choice of those thresholds.

\section*{Data availability}
All data associated with this manuscript are available on Zenodo: \url{https://zenodo.org/records/13837606}.

\section*{Code availability}
All scripts associated with this manuscript are available on Zenodo: \url{https://zenodo.org/records/13837606}.

\section*{Acknowledgements}
L.G. and F.B. acknowledge support from the Air Force Office of Scientific Research under award number FA8655-22-1-7025. 
J.W. acknowledges funding from the Hungarian National Scientific Fund (OTKA FK 145960) and from the European Union under Horizon EU project LearnData (101086712).

\section*{Author Contributions}
F.B. conceptualized the work. 
L.B. collected data and performed the analyses. 
L.G. and J.W. provided methodological insights. 
F.B. supervised the work. 
All authors contributed to the interpretation of the results and participated in the writing of the manuscript.


%

\pagebreak
\clearpage
\widetext
\begin{center}
\textbf{\large Supplementary Material: \\
\mytitle}
\end{center}
\setcounter{equation}{0}
\renewcommand{\theequation}{S\arabic{equation}}
\setcounter{figure}{0}
\renewcommand{\thefigure}{S\arabic{figure}}
\setcounter{table}{0}
\renewcommand{\thetable}{S\arabic{table}}

\setcounter{section}{0}

\setcounter{subsection}{0}

\renewcommand{\thesection}{S\arabic{section}}
\renewcommand{\thesubsection}{\arabic{section}.\arabic{subsection}}

\makeatletter
\def\p@subsection{S}
\makeatother


\begin{figure*}[h!]
    \centering
    \includegraphics[width=\linewidth]{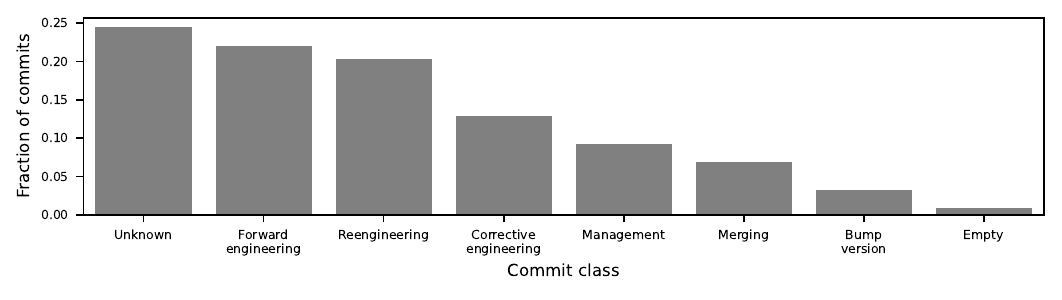}
    \caption{
    Distribution of commit types in the Rust dataset.
    Commits can implement different types of changes to the software and we can use commit messages as an indicator of the type of change being implemented.
    After standard text cleaning (removing file paths, URLs, email addresses, and words shorter than three characters, as well as applying stemming), we applied a keyword-based classifier on commit messages inspired by \citet{hattori2008nature} to label commits as: forward engineering (e.g., ``implement'', ``create''), reengineering (e.g., ``optimize'', ``refactor''), corrective engineering (e.g., ``fix'', ``bug''), and management (e.g., ``clean'', ``documentation'').
    To account for conventions common in GitHub and Rust, we introduced two additional labels: ``bump version'' (Rust-specific jargon indicating the release of a new version of a package) and ``merger'' (merging a pull request or branch).
    In addition, we label as ``unknown'' the commits that do not match any of these six classes and ``empty'' commits whose message is empty.
    Among commits with an identifiable label, 74\% relate to coding tasks (forward, reengineering, corrective) and 26\% to management (merger, management, bump version).
    }
    \label{Sfig:distrib_commit_types}
\end{figure*}

\begin{figure*}[h!]
    \centering
    \includegraphics[width=\textwidth]{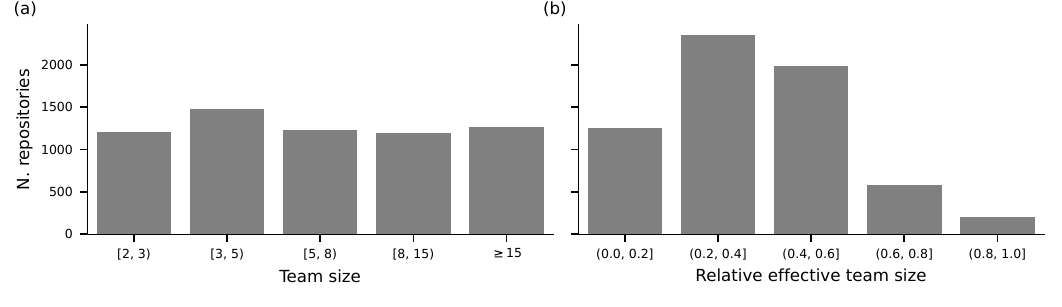}
    \caption{Distribution of repositories across team size (\textbf{a}) and relative effective team size strata (\textbf{b}), as defined in Fig.~\ref{fig:heterogeneity_work_distrib}. 
    }
    \label{Sfig:distribution_team_size_and_releff_team_size}
\end{figure*}

\begin{figure}[h!]
    \centering
    \includegraphics[width=\linewidth]{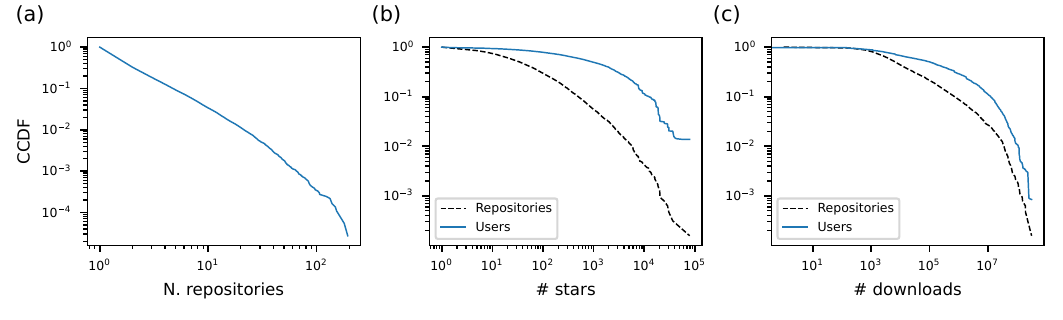}
    \caption{
    \textbf{Distribution of developers' experience and success.}
    Complementary cumulative distribution function (CCDF) of
    (\textbf{a}) number of repositories in which developers make at least one commit, 
    and developers' success in terms of (\textbf{b}) stars and (\textbf{c}) downloads.
    Developers' success is defined as the average number of stars and downloads of repositories where developers make contributions.
    Dashed lines in \textbf{b-c} refer to the CCDF for repositories.  }
    \label{Sfig:developers_experience_and_success}
\end{figure}

\begin{figure}
    \centering
    \includegraphics[width=\linewidth]{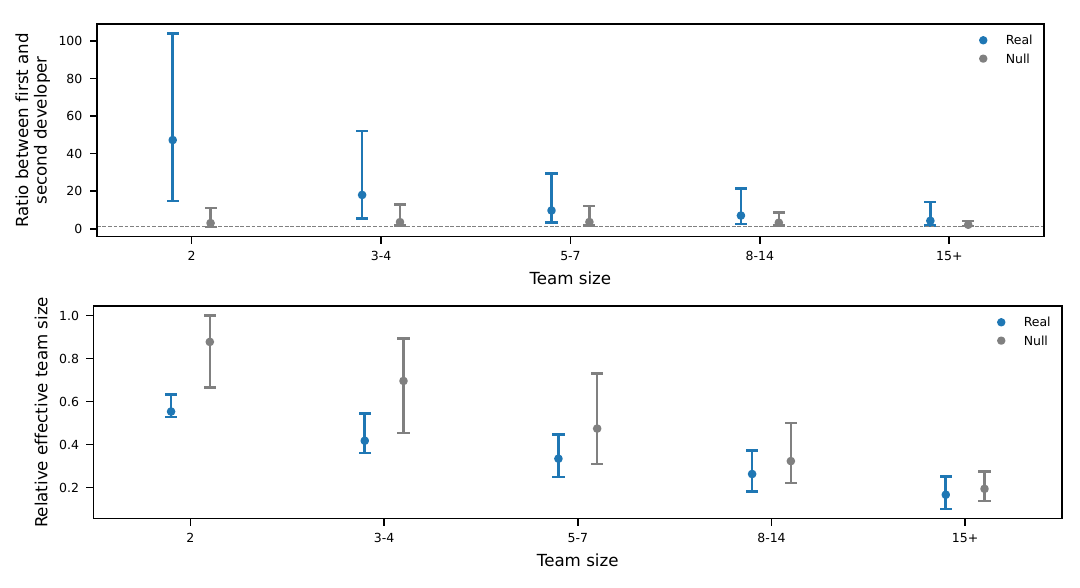}
    \caption{
    Heterogeneity of workload distribution compared to a randomized model.
    We computed the number of commits users makes in each repository, and then shuffle the repositories to which these commits are assigned. 
    This randomization corresponds to a scenario where developers carry a fixed ``effort budget'' (i.e., a fixed number of commits per project) but select repositories at random. 
    The figures show the distribution of quantities in the real data (blue) and in 100 realizations of the randomized model (grey) with medians and error bars ranging from the 25th to the 75th percentile of the distributions.
    (\textbf{a}) Ratio between the number of commits authored by the two most active developers.
    The most active developer contributes substantially more than the second most active developer in the real data compared to the randomized models. 
    (\textbf{b}) Relative effective team size.
    Teams consistently exhibit smaller relative effective team sizes in the real data than in the randomized models, indicating stronger workload heterogeneity. 
    While the differences between the real data and randomized models diminish with larger team sizes, these are all statistically significant (one-sided Mann-Whitney U tests results in $p < 0.001$ for all team sizes). 
    These findings indicate that the observed workload heterogeneity cannot be attributed solely to random allocation of developers' effort.
    }
    \label{Sfig:null_heterogeneity_workload_distribution}
\end{figure}

\begin{figure*}[h!]
    \centering
    \includegraphics[width=\textwidth]{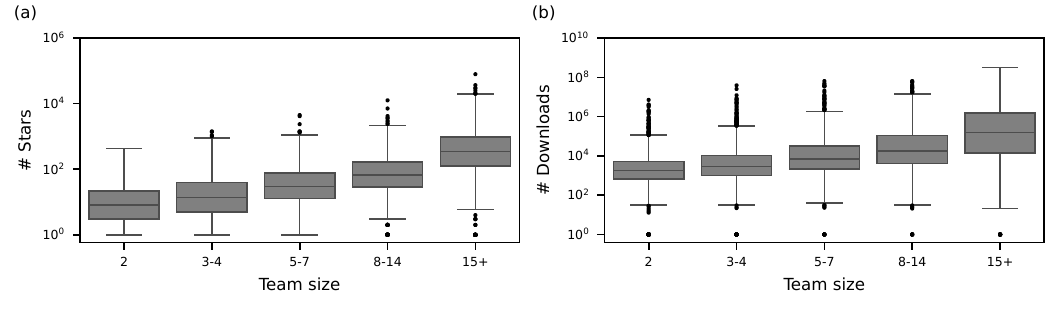}
    \caption{Distribution of number of stars (\textbf{a}) and number of downloads (\textbf{b}) of repositories for different team size strata. 
    The positive correlation is supported by Spearman’s correlation test: $\rho=0.65$ ($p < 0.0001$) for stars and $\rho=0.50$ ($p < 0.0001$) for downloads.
    Box plots indicate median (middle line), 25th, 75th percentile (box) and 5th and 95th percentile (whiskers) as well as outliers (single points).
    }
    \label{Sfig:corr_team_size_success}
\end{figure*}

\begin{figure*}[h!]
    \centering
    \includegraphics[width=\textwidth]{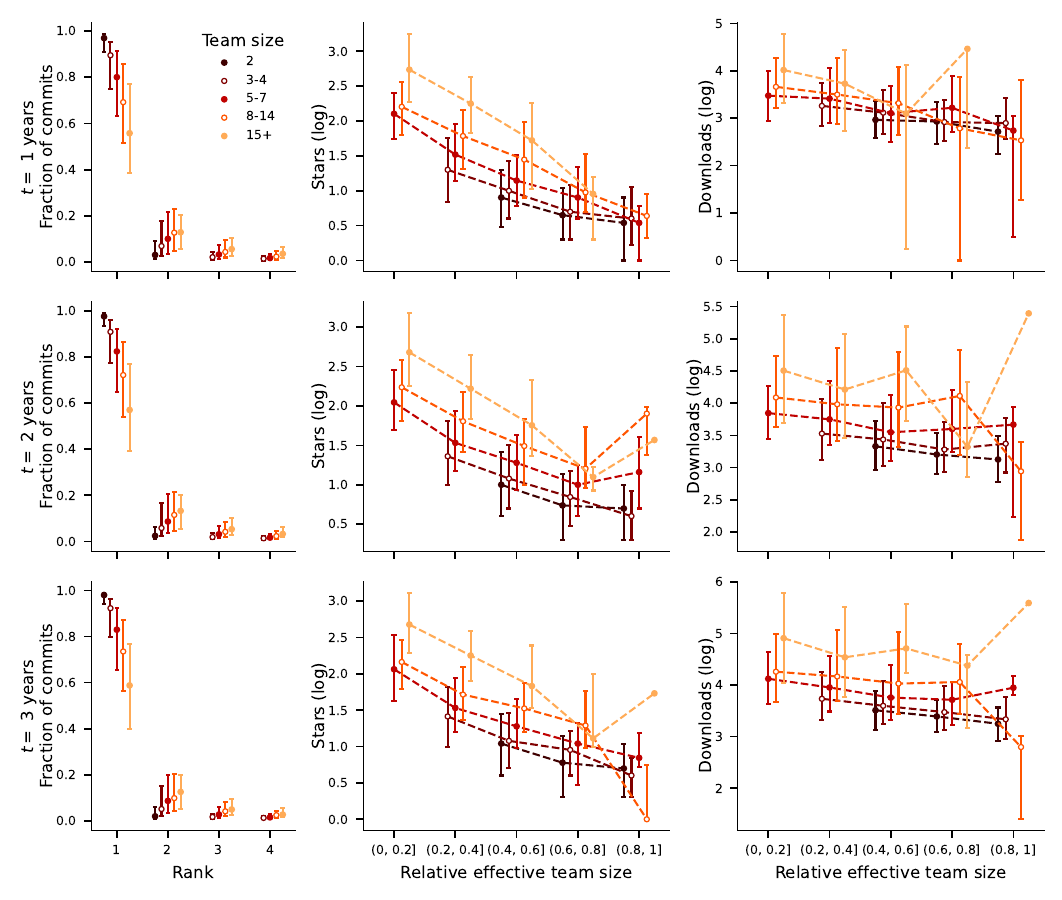}
    \caption{Workload distribution within teams and relationship with success at different point in time during the lifetime of repositories. 
    Similarly to \FigRef{fig:heterogeneity_work_distrib}, the workload distribution is heterogeneous and the success is decreasing in the relative effective team size, pointing to those properties being stable across the lifetime of repositories.
    The Spearman's correlations between relative effective team size and success remains negative and significant for both stars and downloads.
    The only exceptions are downloads for team sizes 8-14 at $t=2$ and $t=3$ years, cases in which the correlation is not anymore significant at 0.05 level.
    Table~\ref{tab:corr_success_rets} shows the detailed results of the correlation.
    The details of the figure are the same as in \FigRef{fig:heterogeneity_work_distrib}.
    }
    \label{Sfig:heterogeneity_work_distrib_year}
\end{figure*}

\setlength{\tabcolsep}{12pt}
\begin{table}[h]
    \centering
    \caption{Spearman's correlation coefficient between the relative effective team size of teams and the success of repositories (stars and downloads) stratified by team size. 
    Correlations are estimated at various stages of the repositories' lifetimes: the last active period in the dataset (All) and after one, two, and three years of activity.    
    Significance levels are denoted as follows:
    * for $p < 0.05$, ** for $p < 0.01$, and *** for $p < 0.001$.
    }
    \begin{tabular}{lllll}
          &     & {} & \multicolumn{2}{c}{Spearman's $\rho$} \\
          &     &  &     Stars & Downloads \\
        Success at & Team size & N. repos. &           &           \\
        \hline
        \multirow{5}{*}{All} & 2 & \num{1212} &   -0.28***  &  -0.16*** \\
          & 3-4 &  1475 &  -0.37*** &     -0.18*** \\
          & 5-7 &  1229 &  -0.41*** &     -0.18*** \\
          & 8-14 &  1192 &  -0.41*** &     -0.11*** \\
          & 15+ & 1263  &  -0.54*** &     -0.16*** \\
          \hline
        \hline
        \multirow{5}{*}{1} & 2 & 1563 &  -0.27*** &  -0.10*** \\
          & 3-4 & 1690 &  -0.37*** &  -0.17*** \\
          & 5-7 & 952  &  -0.42*** &  -0.14*** \\
          & 8-14 & 690  &  -0.42*** &   -0.12** \\
          & 15+ & 372  &  -0.49*** &   -0.15** \\
        \cline{1-5}
        \multirow{5}{*}{2} & 2 & 1354 &  -0.26*** &  -0.11*** \\
          & 3-4 & 1630 &  -0.37*** &  -0.14*** \\
          & 5-7 & 1220 &  -0.42*** &  -0.14*** \\
          & 8-14 & 949  &  -0.44*** &     -0.06 \\
          & 15+ & 728  &  -0.48*** &  -0.13*** \\
        \cline{1-5}
        \multirow{5}{*}{3} & 2 & 929  &  -0.27*** &   -0.10** \\
          & 3-4 & 1229 &  -0.35*** &  -0.15*** \\
          & 5-7 & 1009 &  -0.41*** &  -0.14*** \\
          & 8-14 & 889  &  -0.40*** &     -0.06 \\
          & 15+ & 834  &  -0.48*** &  -0.13*** \\
    \end{tabular}
    \label{tab:corr_success_rets}

\end{table}

\begin{figure*}
    \includegraphics[width=\textwidth]{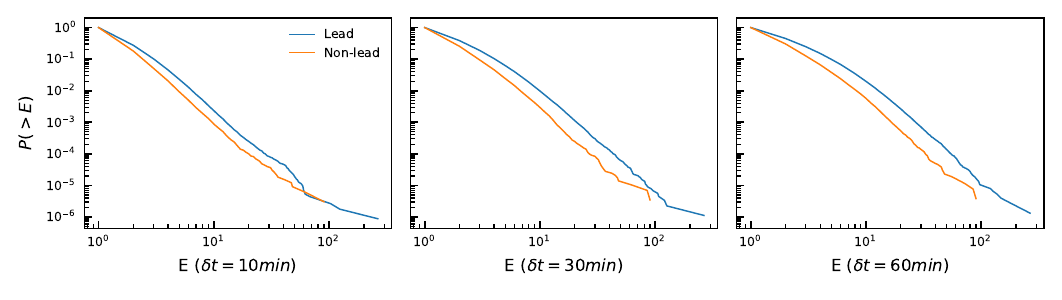}
    \caption{Complementary cumulative distribution of the number of commits close in time $E$ for different values of $\delta t$.
    Leaders are more likely to make commits in larger event train sizes for all the tested values of $\delta t$. 
    }
    \label{Sfig:ccdf_train_events}
\end{figure*}

\begin{figure*}
    \includegraphics[width=\textwidth]{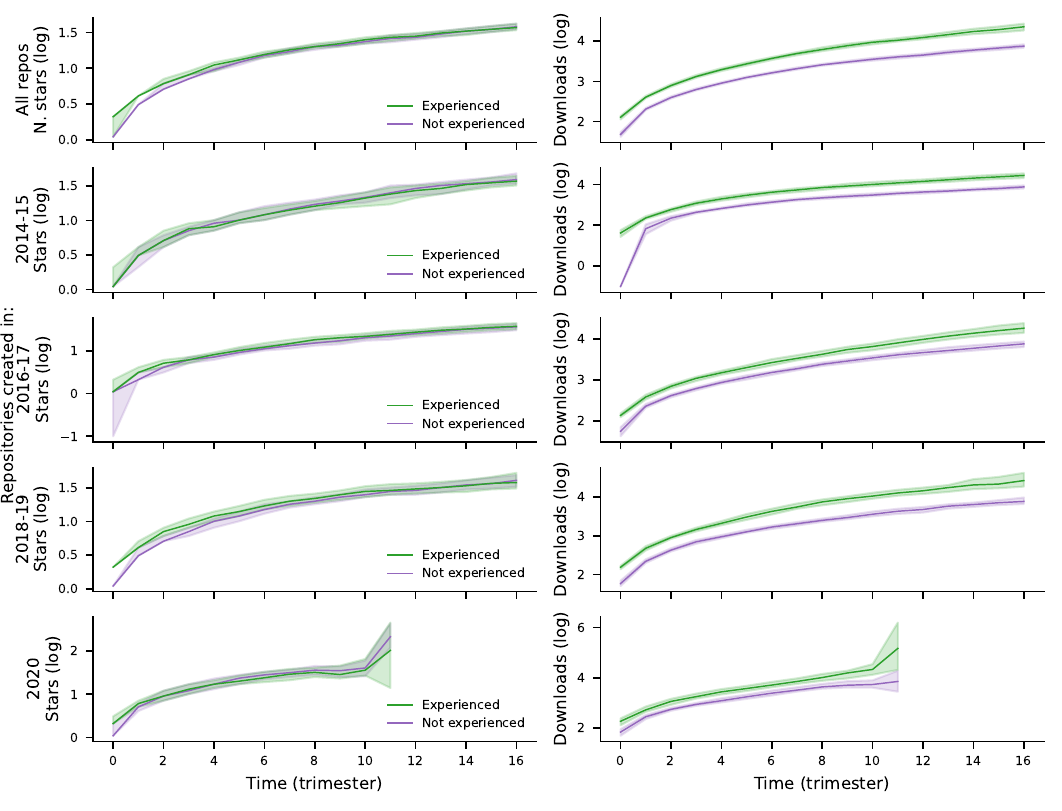}
    \caption{Number of downloads acquired across repositories' lifetime stratified by lead developer's experience (median and its 95\% confidence interval).
    Each row corresponds to repositories initiated in different years, with the first row including all repositories.
    While the prior experience of lead developers doesn't impact repositories' success in terms of stars, it does influence the number of downloads. 
    Indeed, repositories whose lead developer is experienced receive more downloads across their lifetime than those whose lead developer is not experienced.
    After accounting for the initiation year of repositories, we find that this observation remains consistent regardless of the repositories' age.
    }
    \label{Sfig:success_vs_time}
\end{figure*}

\begin{figure*}
    \includegraphics[width=\textwidth]{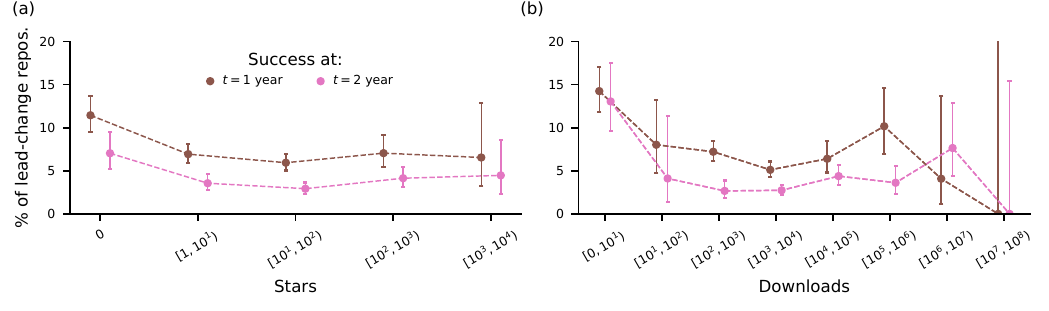}
    \caption{Percentages of repositories that will change the lead developer after time $t$ as a function of the success of repositories at time $t$.
    The figure suggests that repositories with the worst performance in terms of stars (\textbf{a}) and downloads (\textbf{b}) are the most likely to change their lead developer.
    We investigated whether there is a connection between future leadership changes and past success, testing both linear and quadratic dependencies using a fixed effect regression model. 
    However, our analysis did not find significant evidence supporting such a relationship.
    Error bars refer to 95\% confidence intervals of the estimated percentages (Wilson score interval).
    }
    \label{Sfig:leader_change_vs_prev_success}
\end{figure*}

\clearpage

\section{Additional characterization of the activity of lead developers}

\subsection{Lead developers have write access}\label{app:write_access}

In the main manuscript, we identified each developer’s contribution based on the \textit{author} of the commits.
However, a commit may also include a distinct \textit{committer} -- the individual who actually applies and pushes the author’s changes to the repository~\cite{chacon2014git}.
The committer, by definition, has write privileges, whereas the author may not.
We therefore verified whether lead developers appear as committers in the repositories they lead as a proxy of them having write privileges in the projects they lead. 
Our results indicate that in all but six cases, lead developers do indeed have write access to their projects.
Fig.~\ref{Sfig:frac_committers_and_mergers}a shows the fraction of lead and non-lead developers that appear among the committers, stratified for different team sizes.

\subsection{Lead developers are responsible for merging pull requests and branches}\label{app:mergers}

Another important management activity linked to leadership is the process of merging external contributions (pull requests) and integrating separate branches.
Developers involved in these activities have a distinct role within the project that goes beyond having write access and indicate an aspect of leadership.
Here, we quantify to what extent lead developers are involved in these management activities.

\begin{figure}[h!]
    \centering
    \includegraphics[width=\linewidth]{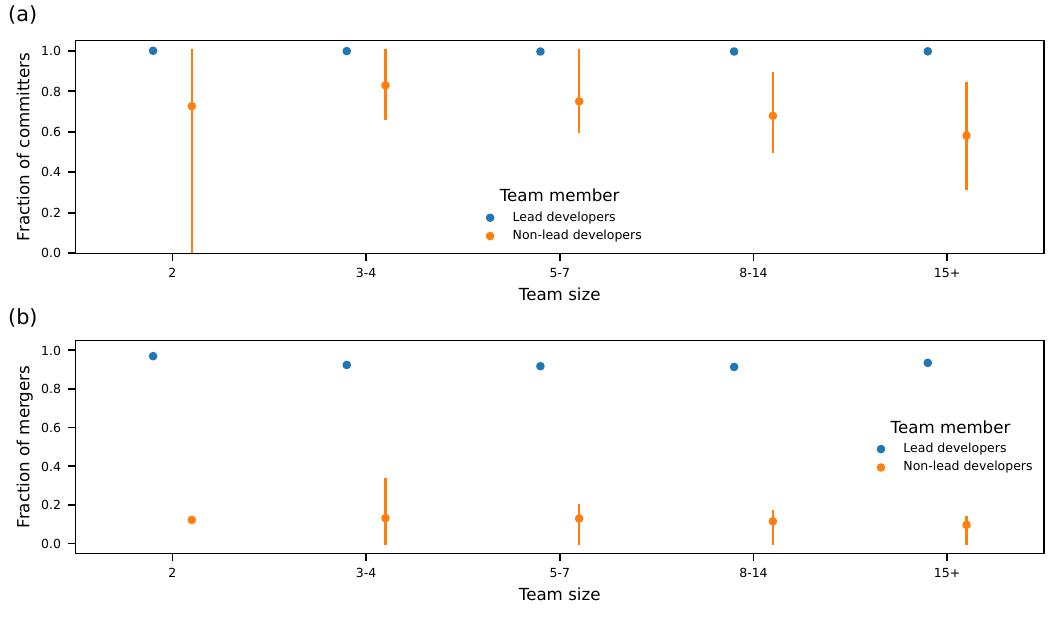}
    \caption{
    \textbf{Involvement of lead developers and non-lead developers in different activities.}
    (\textbf{a}) Fraction of developers among the committers of a repository.
    (\textbf{b}) Fraction of developers among the mergers of a repository.
    Dots indicate fractions averaged across repositories and error bars range between the 25th and 75th percentile of the distributions.
    }
    \label{Sfig:frac_committers_and_mergers}
\end{figure}

\indent First, we identified the commits whose message contains either the string ``\textit{merge pull}'' or ``\textit{merge branch}'', indicating that the commit involves a merging activity.
Then, for each repository, we identified all users involved in merging activities and checked whether the lead developer is among them.
We found that in 95\% of repositories, the lead developer participates in merging activities (83\% if we also count 755 repositories with no identified merge-related commits). 
Moreover, in 87\% of repositories, the developer with the largest number of merge-related commits corresponds to the lead developer (76\% if including the 755 repositories with no identified merge-related commits).
This result suggests that in the vast majority of the repositories, the lead developer is heavily involved in management activities such as merging pull requests or branches.
Figure~\ref{Sfig:frac_committers_and_mergers}b shows the fraction of lead and non-lead developers that appear among the mergers, stratified for different team sizes.

\subsection{Lead developers' position in communication channels}\label{app:coordination_leads}

Issues and pull requests are GitHub's primary coordination tools, allowing developers to discuss and organize tasks, bugs, and plan the implementation of new features.
Here, we analyze the position of the lead developer in the communication network built from the discussions under issues and pull request.
If lead developers occupy leadership roles, we would expect them to be central in the communication network built from these discussion threads.

We constructed a communication network for each repository by considering discussion threads under issues and pull requests.
Specifically, we add a directed link from user $i$ to user $j$ whenever $j$ replies to $i$, thus reflecting the flow of information.
Link weights represent the number of times a directed link is present.
To quantify the importance of a developer in the communication network, we computed the betweenness centrality of nodes, defined as the fraction of shortest paths passing through a specific node.
This has been used in the literature as a measure of coordination in communication networks.

We analyzed separately lead-change and lead-remain repositories.
For lead-remain repositories, we built a single communication network spanning the entire project history. 
We found that lead developers have the largest betweenness centrality in 75\% of the projects and are among the top three in 82\% of the repositories.
This indicates that lead developers are often key coordinators in issue and pull request discussions.

In lead-change repositories, we constructed two separate communication networks: one for the time period in which the old lead developer was in charge (Before LC), and one for the period under the new lead developer (After LC).
Among a subset of these repositories (N=226), only the After LC period showed interactions in issues and pull requests.
Limited to that period, the new lead developer ranks highest in betweenness centrality in 58\% of cases (and in the top three in 86\%), while the old lead developer ranks highest in only 8\% of cases (top three in 38\%).
Thus, once the new lead developer takes over, they frequently occupy a highly central position in the communication network as compared to the old lead developer.

\begin{figure}
    \centering
    \includegraphics[width=0.99\linewidth]{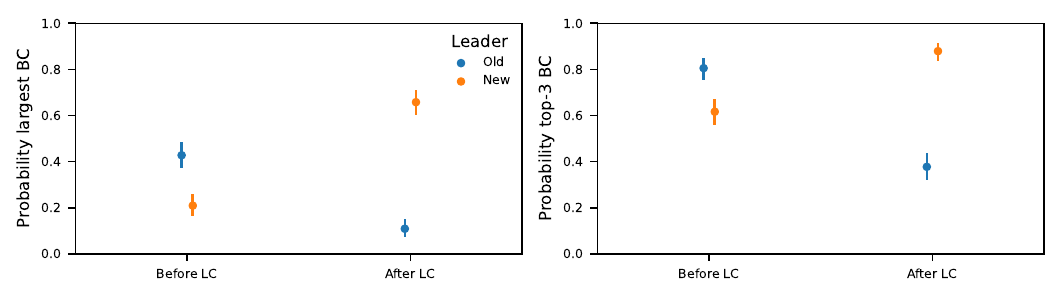}
    \caption{
    \textbf{Centrality of lead developers in repositories' coordination activities.} 
    The figure refers to lead-change repositories with activity in both before and after periods.
    (\textbf{a}) Fraction of repositories where old (blue) and new lead developers (orange) have the largest betweenness centrality. 
    (\textbf{b}) Fraction of repositories where where the two lead developers are among the top three developers in terms of betweenness centrality.
    Before LC and After LC refers to the time span when the old and the new lead developers were in charge, respectively.
    Error bars refer to 95\% confidence intervals of the estimated fractions (Wilson score interval).
    }
    \label{R1:compare_betweenness_before_after}
\end{figure}

\begin{figure*}
    \includegraphics[]{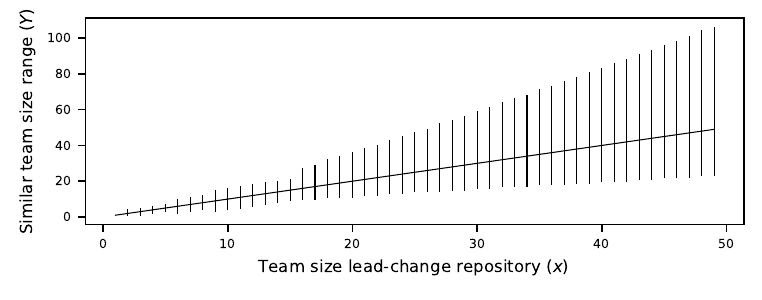}
    \caption{Graphical representation of the team size similarity used in the matching approach.
    For each team size of the lead-change repository (horizontal axis), the range of similar team sizes for the candidate repositories is represented (vertical lines).
    The solid line represents the two team sizes being equal.
    In plain terms, we require similar team sizes to be strictly close or equal for small team sizes (2 to 15), while being similar in relative terms for larger team sizes (larger than 15).
    The functional relationship is the following, considering $x$ being the team size of the lead change repository and $Y$ being the range of similar team sizes for the candidate repositories: 
    $Y=\left\{1\right\}$ if $x=1$, 
    $Y=\left\{n | n \in \mathbb{N} \text{ and } |n - x| \leq 2\right\}$ if $x \in \left[2, 5 \right]$, 
    $Y=\left\{n | n \in \mathbb{N} \text{ and } |n - x| \leq 4\right\}$ if $x \in \left[6, 8 \right]$, 
    $Y=\left\{n | n \in \mathbb{N} \text{ and } |n - x| \leq 6\right\}$ if $x \in \left[6, 15 \right]$,
    $Y=\left\{n | n \in \mathbb{N} \text{ and } |\log_{10}(n) - \log_{10}(x)| / \log_{10}(x) \leq 0.2\right\}$ if $x > 15$.    
    }
\label{Sfig:similar_team_sizes_match}
\end{figure*}

\begin{figure*}
    \includegraphics[width=\textwidth]{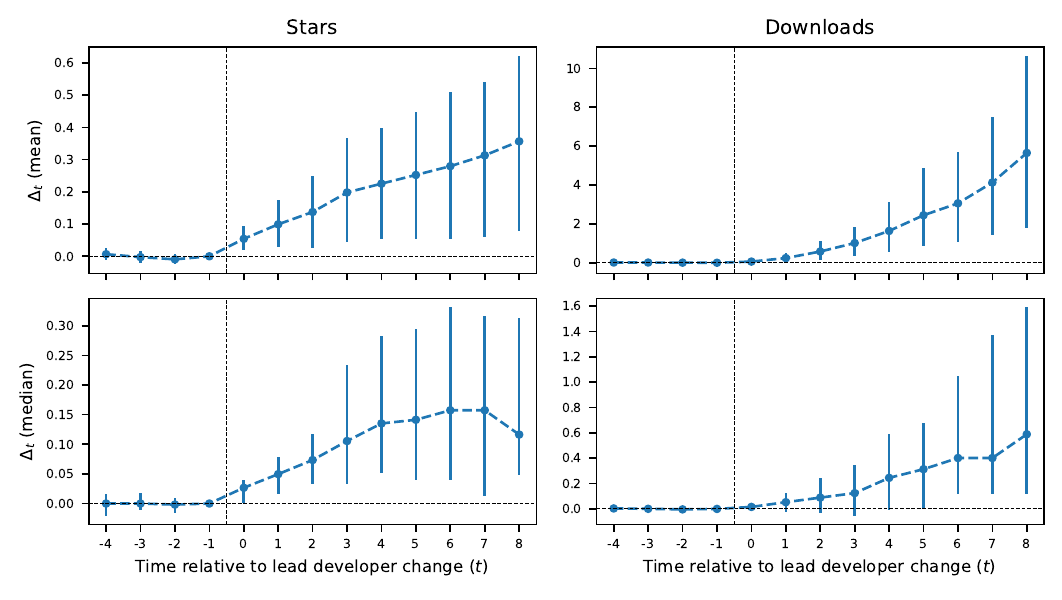}
    \caption{We checked if the result for the average $\Delta_t$ (top row) is influenced by the skewness of its distribution by comparing it with the estimation of the median $\Delta_t$ (bottom row).
    The effect on stars (left column) and downloads (right column) remains positive, although the magnitude decreases, especially for downloads.
    This suggests that while the skewness of $\Delta_t$ may enhance the average $\Delta_t$, it doesn't affect the direction of the effect. 
    Error bars represent 95\% confidence intervals of the estimated quantities.
    }
\label{Sfig:effect_leader_change_mean_median}
\end{figure*}

\begin{figure*}
    \includegraphics[width=\textwidth]{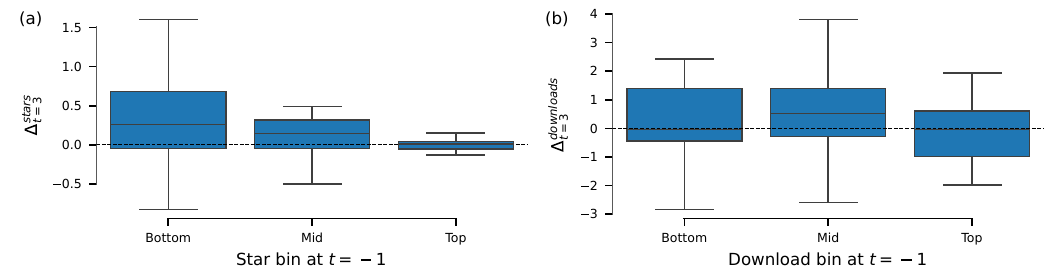}
    \caption{Descriptive analysis of the heterogeneity of $\Delta_t$ considering the success before the change of the lead developer.
    Repositories are divided into bins depending on the number of stars (\textbf{a}) and number of downloads (\textbf{b}) in the trimester preceding the change of the lead developer ($t=-1$). 
    Bottom teams: below the 30th percentile of success; Mid teams: between the 50th and 80th percentile of success; Top teams: above the 90th percentile of success.
    Although worst performing repositories in terms of stars benefit more from the change of lead developer (Kendall's $\tau_b=-0.18$, $p = 0.03$), this is not the same when we consider downloads.
    Indeed, we observe no difference on the effect depending on the number of downloads before the change of the lead developer (Kendall's $\tau_b=-0.01$, $p = 0.92$).
    Box plots indicate median (middle line), 25th, 75th percentile (box) and 5th and 95th percentile (whiskers).
    }
    \label{Sfig:het_effect_previous_success}
\end{figure*}

\begin{figure*}
    \includegraphics[]{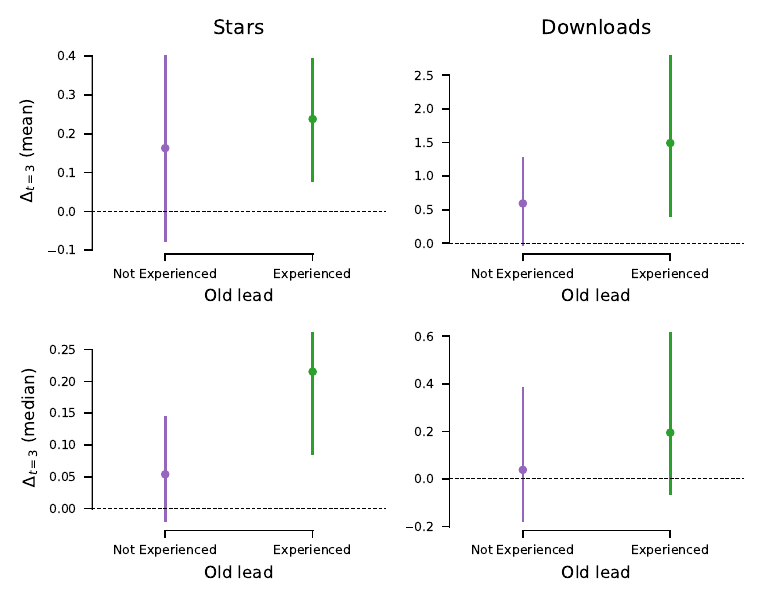}
    \caption{Descriptive analysis of the heterogeneity of $\Delta_t$ at $t=3$ considering the experience of the old lead developer.
    Repositories initiated by experienced lead developers have a benefit from the turnover (stars: $T=1488$, $n=63$, $p=0.0004$, one sided; downloads: $T=1314$, $n=63$, $p=0.02$, one sided), whereas repositories initiated by unexperienced developers have a significant benefit only in terms of stars (stars: Not experienced: $T=1682$, $n=72$, $p=0.02$, one sided; downloads: $T=1488$, $n=63$, $p=0.0004$, one sided).
    The second row displays the median $\Delta_t$, indicating that the observed result is not affected by the skewness of the distribution.
    Error bars refer to 95\% confidence intervals of the estimated quantities.
    }
    \label{Sfig:het_effect_leader_experience}
\end{figure*}

\begin{figure*}
    \includegraphics[]{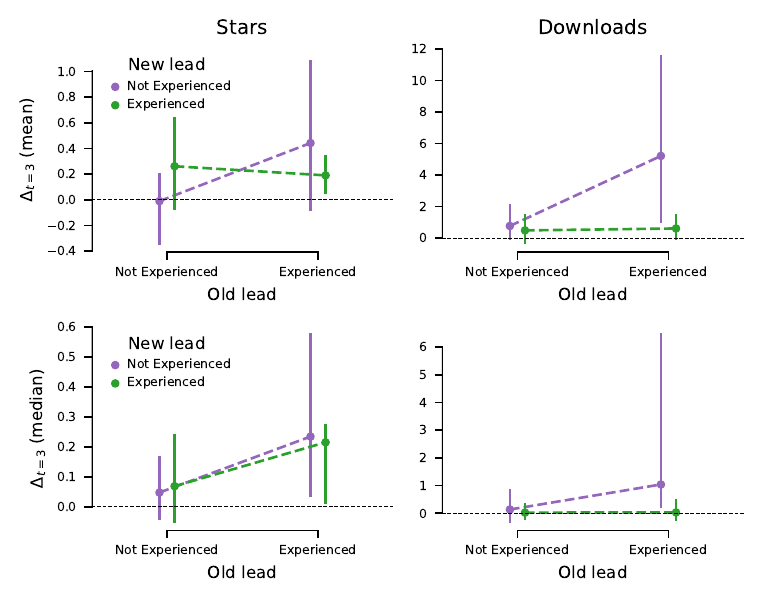}
    \caption{Descriptive analysis of the heterogeneity of $\Delta_t$ considering the experience of the old and new lead developer.
    The figure suggests that the experience of the new lead developer is not affecting strongly the way repositories benefit from the change.
    However, this further stratification shrinks the sample size and the large error bars make it difficult to draw conclusions from this analysis.
    The second row displays the median $\Delta_t$.
    Error bars refer to 95\% confidence intervals of the estimated quantities.
    }
    \label{Sfig:het_effect_both_leader_experience}
\end{figure*}

For the remaining lead-change repositories -- those displaying activity in both time periods -- we compared the old and new lead developers’ betweenness centrality ranks before and after the leadership change, and show the result in Fig.~\ref{R1:compare_betweenness_before_after}a.
Prior to the change, the old lead developer had the highest betweenness in more than 40\% of repositories (over twice as often as the new lead developer), and together they appeared as top-ranked in over 60\% of cases.
When examining the top three developers in terms of betweenness centrality (Fig.~\ref{R1:compare_betweenness_before_after}b), the old and new leads appeared in 80\% and 60\% of repositories, respectively.
After the change, the new lead developer had the highest betweenness in more than 60\% of repositories, while the rank of the old lead developer drops dramatically. 
These results points to lead developers being important actors in the communication network and in coordinating activities within the project.

One caveat is that the smaller gap between old and new lead developers in the before period may reflect an artifact of our identification of lead developer change.
Indeed, we identify the transition as the point at which the new lead surpasses the old lead in terms of number of commits. 
In practice, the transition likely begins earlier, as the amount of new lead’s contributions gradually increases while the one of the old lead declines (see Figure~3b in the main manuscript).

\clearpage

\onecolumngrid

\begin{table*}[!htbp]
    \caption{
    We used a fixed-effect regression model to verify that our finding on the relationship between lead developer change and success is not confounded by influence of the lead developer (e.g., lead developers with different skills, popularity).
    This design allows us to separate the effect of transitioning from one lead developer to another from any unobserved heterogeneity associated to lead developers.
    The predictors include all the features used for the matching: is\_lead\_change\_t, indicating whether the old or new lead developer is in charge, team\_size\_t, being the log-transformed team size, and rel\_eff\_team\_size\_t, being the relative effective team size.
    Features are measured at time $t$.
    In addition to the lead developers' fixed effects (lead\_t), we included fixed effects for repositories (i.e., repo\_id) and for the date of trimester $t$ (period\_month), which capture repositories' unobserved heterogeneity and system-level shocks. 
    Models (1) and (3) align with the result of the matching presented in the main text.
    Indeed, the estimated coefficient for the change of lead developer is positive and statistically significant (0.2361 and 0.3593, respectively).
    Models (2) and (4) return similar results (0.1166 and 0.3817, respectively), thus confirming that the observed association between lead developer change and success is not confounded by the influence of lead developers.
    }
    \label{Sfig:panel_regression_leader_fe}
    \centering
    \renewcommand{\arraystretch}{1.2} 
    \setlength{\tabcolsep}{8pt} 
    \begin{tabular}{lcccc}
       \hline \hline
       Dependent Variables: & \multicolumn{2}{c}{$\log_{10}(\# stars)$} & \multicolumn{2}{c}{$\log_{10}(\# downloads)$}\\
       Model:                      & (1)             & (2)             & (3)            & (4)\\  
       \hline
       \emph{Variables}\\
       is\_lead\_change\_t         & 0.2361$^{***}$  & 0.1166$^{**}$   & 0.3593$^{***}$ & 0.3817$^{**}$\\   
                                   & (0.0101)        & (0.0517)        & (0.0316)       & (0.1801)\\   
       team\_size\_t               & 0.0039$^{***}$  & 0.0037$^{***}$  & 0.0075$^{***}$ & 0.0079$^{***}$\\   
                                   & (0.0001)        & (0.0002)        & (0.0003)       & (0.0003)\\   
       rel\_eff\_team\_size\_t     & -0.9493$^{***}$ & -0.9461$^{***}$ & -1.689$^{***}$ & -1.693$^{***}$\\   
                                   & (0.0247)        & (0.0249)        & (0.0546)       & (0.0537)\\   
       \hline
       \emph{Fixed-effects}\\
       repo\_id                    & Yes             & Yes             & Yes            & Yes\\  
       period\_month               & Yes             & Yes             & Yes            & Yes\\  
       lead\_t                   &                 & Yes             &                & Yes\\  
       \hline
       \emph{Fit statistics}\\
       Observations                & 116,296         & 116,296         & 116,296        & 116,296\\  
       R$^2$                       & 0.92492         & 0.93162         & 0.86749        & 0.87598\\  
       Within R$^2$                & 0.27083         & 0.26011         & 0.18000        & 0.17570\\  
       \hline \hline
       \multicolumn{5}{l}{\emph{Driscoll-Kraay (L=3) standard-errors in parentheses}}\\
       \multicolumn{5}{l}{\emph{Signif. Codes: ***: 0.01, **: 0.05, *: 0.1}}\\
    \end{tabular}
\end{table*}

\clearpage

\section{Lead developers: a more restrictive definition}\label{app:alternative_lead_detection}

In the main text, we defined the lead developer of a repository at trimester $t$ as the developer who authored the largest share of commits up to that time.
However, this definition does not account for the fact that the number of commits of the most active developer may not be significantly larger than the one of the second most active developer.
We thus compared our results against an alternative definition of lead developer, which works as follows:
\begin{itemize}
    \item[1.] For each trimester of activity, we count the cumulative number of commits for all developers and let $n_1$ and $n_2$ be the number of commits of the most and second-most active developer, respectively.
    \item[2.] We test whether $n_1$ is significantly larger than $n_2$ using a one-sided binomial test to check if the fraction of commits by the most active developer (i.e., $n_1 / (n_1 + n_2)$) is larger than 0.5 at 5\% significance level.
\end{itemize}
The new definition takes the same input as the original definition of lead developer (i.e., the cumulative commit counts) but introduces a statistical test to determine whether the top developer ``significantly'' leads.
A noteworthy property of this test is that its power grows with $\sqrt{n_1 + n_2}$.
Consequently, the same difference $n_1 - n_2$ can be statistically significant when the overall commit count is low yet become non-significant if the total commit count is large.
For instance, $n_1=25$ and $n_2=5$ is a significant difference (one-sided $\text{p-value} < 0.001 $), whereas $n_1=120$ and $n_2=100$ is not (one-sided $\text{p-value} = 0.10 $).
In other words, the threshold difference $n_1 - n_2$ that is deemed not significant is an increasing function of $n_1 + n_2$.
This feature is desirable because it prevents from identifying a top developer as a ``significant lead'' when the margin is proportionally too small. 

Our new definition naturally introduces situations in which no developer passes the test, giving rise to what we refer to as an ``ambiguous lead developer.''
We handle this situation in the following ways:
\begin{itemize}
    \item[(a)] when ambiguous lead developers are observed in the middle of a transition of lead developers, we assume that the transition occurred in the middle of such time span and fill the ambiguous lead developers accordingly.
    It is reasonable to expect such a situation because of the takeover: one developer that surpasses the previous lead developer in terms of commits, thus there may be a time span where these two developers made a comparable contribution.
    \item[(b)] when ambiguous lead developers occur at any time $t$ and the lead developer is user $i$ at both time $t-1$ and $t+1$, we fill the ambiguous lead at time $t$ to be user $i$.
    We interpret this situation as noise.
\end{itemize}
Following this new definition, we found 555 repositories to have at least one trimester with an ambiguous lead developer, which we discarded.
Among these 555 repositories, 396 are lead-remain and 159 are lead-change repositories.
Notably, the proportion of removed lead-change repositories is higher because those are more prone to having periods where the top two developers’ commit counts are close.

\begin{figure}[h!]
    \centering
    \includegraphics[width=\linewidth]{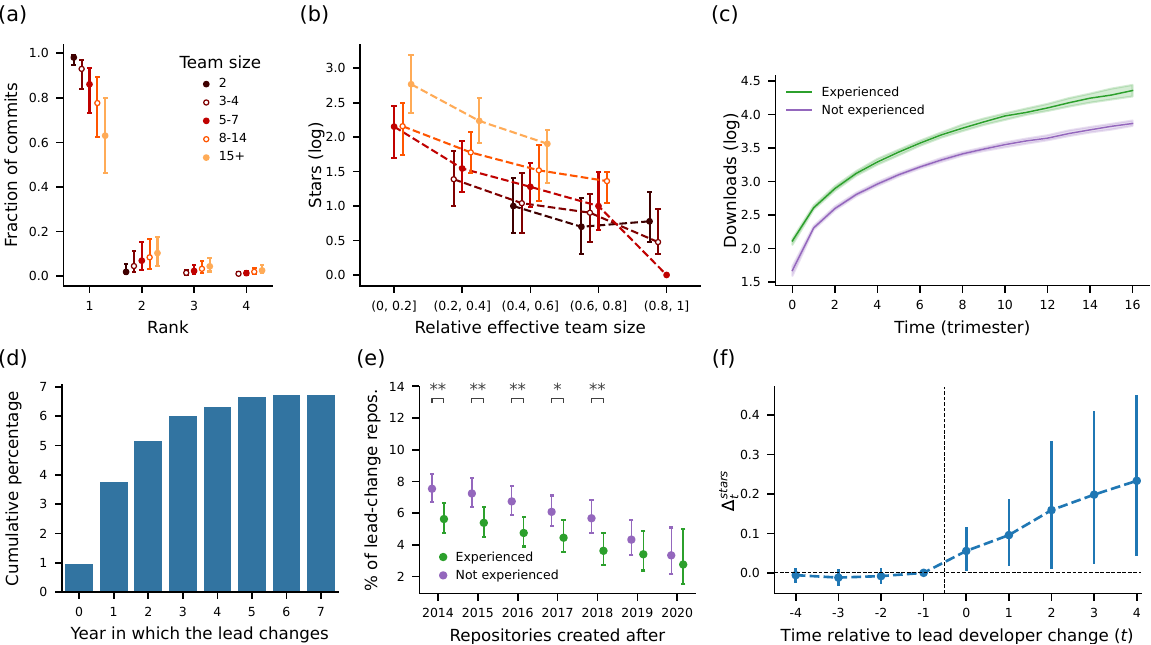}
    \caption{
    Replication of the main results using the more robust definition of lead developers.
    (\textbf{a}) Heterogeneous workload distribution within teams (replicates Fig.~1a).
    (\textbf{b}) Negative correlation between relative effective team size and number of stars (replicates Fig.~1b).
    (\textbf{c}) Number of downloads across repositories’ lifetime stratified by lead developers’ experience (replicates Fig.~2d).
    (\textbf{d}) Cumulative percentage of repositories undergoing a lead developer change as a function of the number of years since their creation (replicates Fig.~3a).
    (\textbf{e}) Percentage of lead-change repositories stratified by the previous experience of the old lead developer (replicates Fig.~3c).
    (\textbf{f}) Average effect of lead developer change $\Delta t$ for stars (replicates Fig.~4a).
    We replicated the results also for all the other analyses not shown in this figure.
    }
    \label{R1:replicate_with_other_lead_definition}
\end{figure}

At this point, we can apply the exact same analyses presented in the main manuscript.
We were able to fully replicate all our findings: (i) the workload distribution is heterogeneous and this property correlates with higher success (Fig.~\ref{R1:replicate_with_other_lead_definition}a-b), (ii) lead developers make commits in longer burst and contribute in a larger number of repositories than non-lead developers, with their experience being associated to more downloads (Fig.~\ref{R1:replicate_with_other_lead_definition}c), (iii) 7\% of repositories change the lead developer at some point, with repositories led by experienced lead developers being less likely to undergo such a transition (Fig.~\ref{R1:replicate_with_other_lead_definition}d-e), and (iv) lead developer change is associated to an increase in success after the takeover (Fig.~\ref{R1:replicate_with_other_lead_definition}f).

\section{Replication of analysis on PyPI and NPM packages}\label{app:replication_analysis}

\subsection{Data collection}\label{app:replication_analysis_data_collect}

We used data sourced from~\cite{marat2018ecosystem} to replicate the analysis of the main paper on two additional datasets.
The first dataset comprises all commits of Python packages that have an associated GitHub repository.
The second dataset, while not discussed in the reference~\cite{marat2018ecosystem}, is included in the paper's repository~\cite{marat2018ecosystemData} and contains commits from JavaScript packages with corresponding GitHub repositories.
In both cases, the repository activity is tracked until January 2018, corresponding to the time of the data collection.
We chose these datasets because they contain information on packages linked to GitHub repositories, similarly to the Rust dataset.
Although not as comprehensive as the Rust dataset, these additional datasets provide a good starting point for the collection of additional information that can allow us to perform the same analyses across the three programming languages.
In the following, we refer to these data sources as the JavaScript dataset and the Python dataset to distinguish them from the Rust dataset.

The Python and JavaScript datasets lack success metrics of repositories, which are crucial for our analyses.
We thus retrieved stars using the GitHub REST API~\footnote{\url{https://docs.github.com/en/rest?api}} and JavaScript package downloads using the NPM REST API~\footnote{\url{https://github.com/npm/registry/blob/main/docs/REGISTRY-API.md}} (NPM is the package registry of JavaScript).
We did not consider download data for Python packages, as these time series start from 2016. 
Additionally, the service that tracked download statistics for PyPI packages experienced frequent disruptions until July 2018~\footnote{\url{https://pypistats.org/faqs#why-are-there-so-many-more-downloads-after-july-26-2018}}.
The downloads time series is incomplete for the JavaScript dataset as well since it starts from the beginning of 2015. 
Therefore, all the analyses involving the download time series of JavaScript repositories are limited to repositories that were created after January 2015 (47\% of the total).

Differently from the Rust dataset, the JavaScript and Python datasets have a commit disambiguation issue, where the same commit can be attributed to multiple repositories.
This duplication arises because forks inherit the history of commits of the forked repositories.
For our analysis, this constitutes a problem because we would count several times the activity of a team.
To solve this issue, we first collected the date of creation of repositories using the GitHub Rest API and disambiguate duplicated commits by assigning them to the older repository.
Although we have no guarantee of assigning the commit to the top repository in the fork tree, we rule out the possibility of overcounting the activity of the same team.
This approach is similar to the one used for the data collection of the Rust dataset by~\cite{schueller2022evolving}.

An additional issue, not encountered in the Rust dataset, is the presence of committers with missing name.
Since this information cannot be recovered, we treated all commits with an unknown committer as authored by the same user.
Then, we discarded repositories where this unknown user was the lead developer at any point in time.
In that way, we discard repositories where a meaningful fraction of commits could not be reliably attributed.

After the collection of these additional information, we created two datasets that are consistent with the Rust dataset.
After applying the same filters for the selection of repositories as described in the Methods section, the total number of Python repositories is \num{10339} and \num{22662} for JavaScript, which are respectively 1.7 and 3.7 times bigger than the number of repositories in the Rust dataset.

\begin{figure*}
	\includegraphics[width=\textwidth]{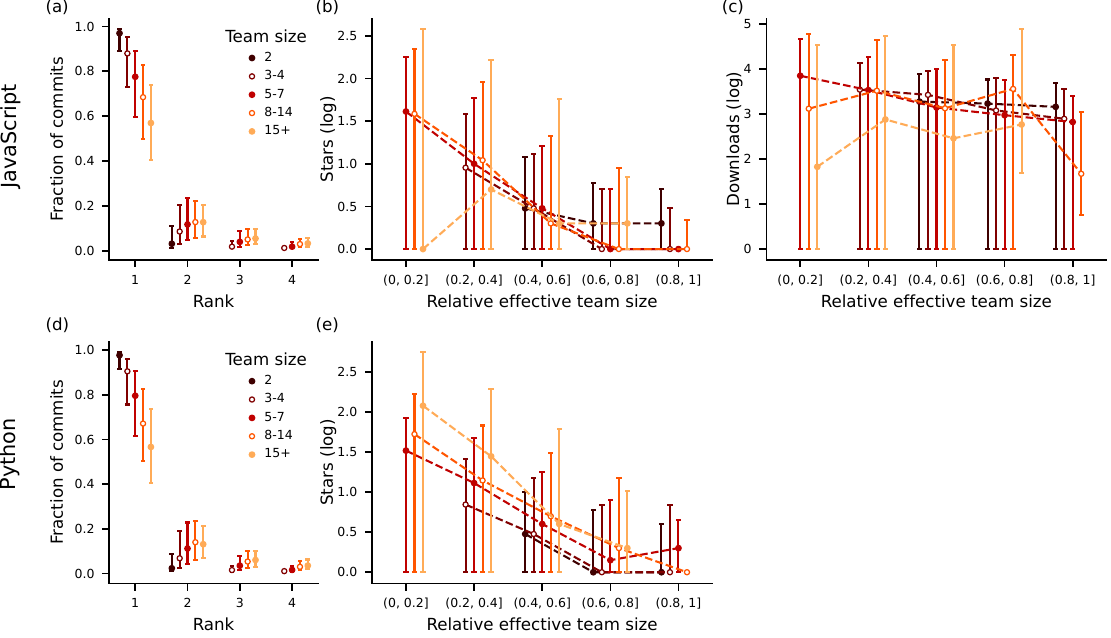}
	\caption{\textbf{Workload distribution within teams and relationship with success.}
		This figure replicates \FigRef{fig:heterogeneity_work_distrib} of the main paper.
		(\textbf{a}, \textbf{d}) The workload distributions resemble the one observed for Rust repositories: one developer makes most of the commits (i.e., the lead developer) and other developers contribute to a lesser extent.
		(\textbf{b-c}, \textbf{e}) The negative correlation between relative effective team size and success is significant (Spearman's rank test $p < 0.05$), indicating that the more heterogeneous the workload distribution in the team, the higher the success.
        The correlation coefficients indicate a weaker correlation compared to Rust.
        Indeed, the Spearman's correlation coefficient for stars ranges from -0.14 to -0.27 in JavaScript and from -0.16 to -0.25 in Python.
        For what concerns downloads, the coefficient ranges from -0.03 to -0.11 in JavaScript.
        The only exception is for downloads in JavaScript for large team size (15+), which returns a positive and significant correlation coefficient of 0.10.
        Error bars are wider than in \FigRef{fig:heterogeneity_work_distrib} because the fraction of repositories with no stars is larger than in Rust: 5\% for Rust, 41\% for JavaScript, and 38\% for Python.
        The larger share of repositories with no stars likely contributes in weakening the correlations between success and heterogeneity of workload distribution.
		Note that we have no data for downloads of Python repositories.
	}
	\label{fig:heterogeneity_work_distrib_js_py}
\end{figure*}

\begin{figure*}[t]
	\includegraphics[width=\textwidth,trim={0cm 2.5cm 0.5cm 0cm},clip]{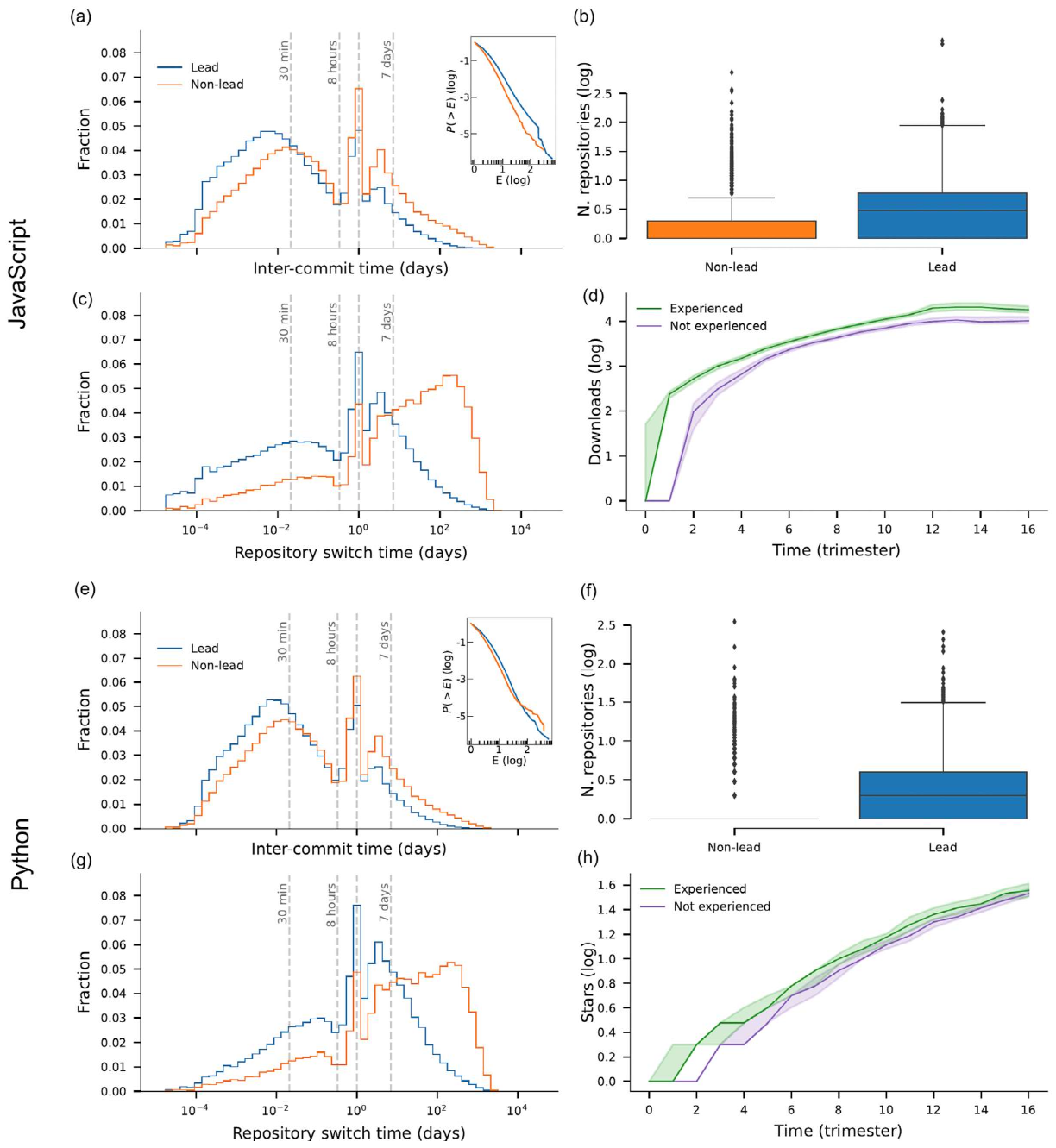}
	\caption{\textbf{Characterization of lead developers' activity.}
		This figure replicates \FigRef{fig:leaders_activity} of the main paper.
		(\textbf{a}, \textbf{e}) The inter-commit time distribution of lead developers is left skewed compared to non-lead developers and have longer streaks of consecutive commits (as shown in the inset), except for Python repositories.
		This discrepancy may be attributed to the presence of bot accounts that are not filtered out in the JavaScript and Python datasets.
		(\textbf{b}, \textbf{f}) Lead developers participate in more repositories compared to non-lead developers.
        The differences are deemed significant according to the Mann–Whitney \textit{U} test (JavaScript: $U=\num{1013358981.5}$, $n_1=\num{11910}$ and $n_2=\num{110035}$, $p<0.0001$; Python: $U=\num{289640558.5}$, $n_1=\num{7069}$ and $n_2=\num{56935}$, $p<0.0001$).
		(\textbf{c}, \textbf{g}) Lead developers tend to switch from one project to another on a daily to weekly basis.
		(\textbf{d}) JavaScript repositories led by experienced developers are downloaded more compared to those led by inexperienced ones.
        (\textbf{h}) Python repositories receive a comparable number of stars independently from the experience of their lead developer.
		Note that we have no data for downloads of Python repositories.
        }
	\label{fig:leaders_activity_js_py}
\end{figure*}

\begin{figure*}[t]
	\includegraphics[width=\textwidth]{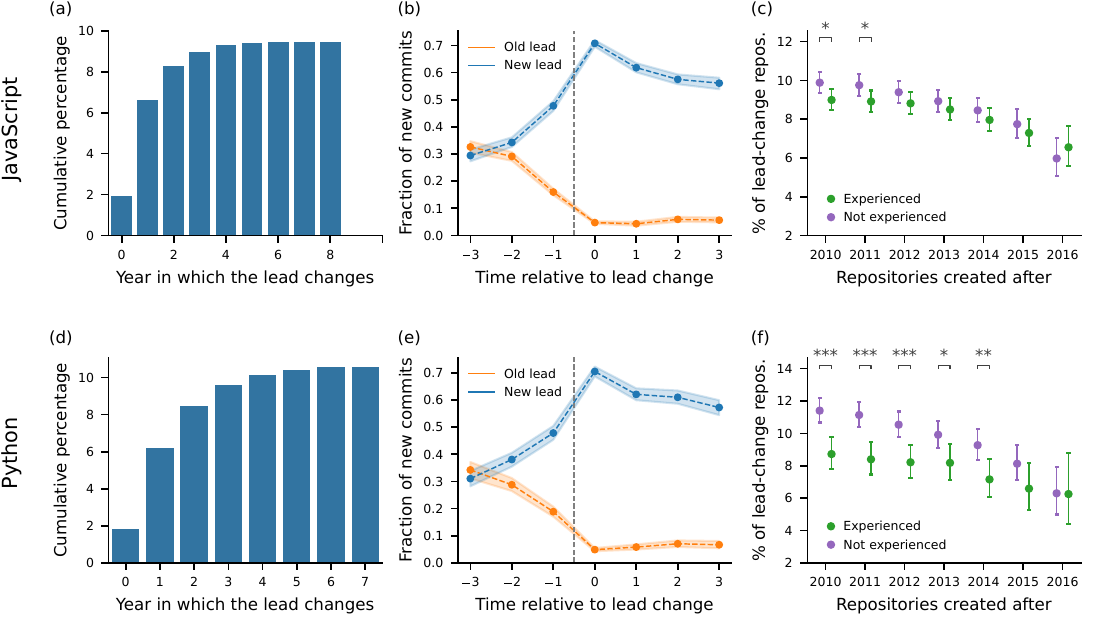}
	\caption{\textbf{Lead developers can change across the lifetime of repositories.}
        This figure replicates \FigRef{fig:leader_changes} of the main paper.
		(\textbf{a}, \textbf{d}) 9\% of JavaScript and 11\% of Python repositories experience a change of lead developer, with the majority occurring within the second and third year of activity.
        These percentages are close to the one observed in Rust, where 10\% of repositories changed their lead developer.
		(\textbf{b}, \textbf{e}) One year before the transition, the old and new lead developer contribute similarly, while after the transition (vertical dashed line), contributions from the old lead developer diminish rapidly.
		(\textbf{c}, \textbf{f}) Repositories led by experienced lead developers are less likely to change the lead developer (odds ratio at $1.11$, $p = 0.02$ for JavaScript, odds ratio at $1.35$, $p < 0.0001$ for Python).
		The strength of the association is weaker for JavaScript repositories.
	}
	\label{fig:leader_changes_js_py}
\end{figure*}

\begin{figure*}[t]
	\includegraphics[width=\textwidth]{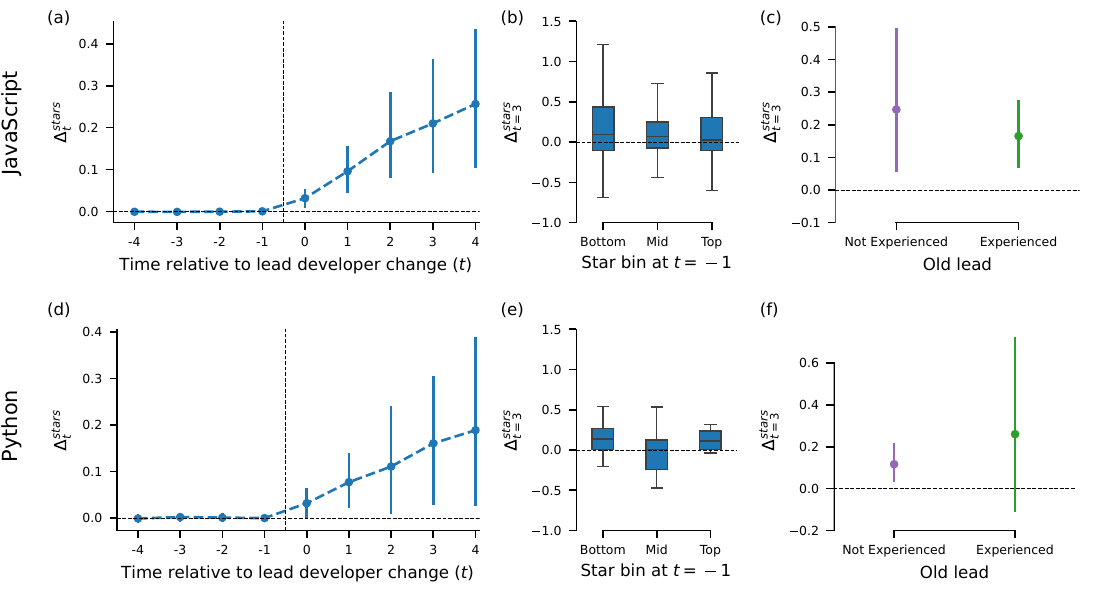}
	\caption{\textbf{Lead developer changes are associated with faster success growth.}
        This figure replicates \FigRef{fig:leader_changes_correlation_success} of the main paper.
		(\textbf{a}, \textbf{d}) Repositories' success grows faster compared to similar repositories that did not undergo such a change, with a magnitude comparable to Rust repositories.
        (\textbf{b-c}, \textbf{e-f}) Differently from Rust, there is no difference in success growth with respect to (\textbf{b}, \textbf{e}) the success before the change of lead developer (JavaScript: Kendall's $\tau_b=-0.12$, $p = 0.17$; Python: Kendall's $\tau_b=-0.05$, $p = 0.47$) and (\textbf{c}, \textbf{f}) the experience of the old lead developer (JavaScript-Not experienced: $T=2906$, $n=94$, $p=0.005$, one sided; JavaScript-Experienced: $T=2143$, $n=78$, $p=0.001$, one sided; Python-Not experienced: $T=2027$, $n=78$, $p=0.007$, one sided; Python-Experienced: $T=401$, $n=35$, $p=0.08$, one sided).
	}
	\label{fig:leader_changes_correlation_success_js_py}
\end{figure*}

In the following, we show in \FigRef{fig:heterogeneity_work_distrib_js_py}, \FigRef{fig:leaders_activity_js_py}, \FigRef{fig:leader_changes_js_py}, and \FigRef{fig:leader_changes_correlation_success_js_py} the results of the analysis on the JavaScript and Python datasets.
All the main results are consistent and in line with the ones of the Rust dataset. 
We discuss the main differences in the captions of the figures.

\end{document}